\documentstyle[prd,aps,epsf,float]{revtex}
\begin{document}

\title{\bf Color superconductivity and its electromagnetic manifestation}
\onecolumn
\draft

\author{S. Ying}
\address{Research Center for Theoretical Physics, 
Physics Department, Fudan University\\
Shanghai 200433, People's Republic of China}
\def\bra#1{\mathopen{\langle#1\,|}}
\def\ket#1{\mathclose{|\,#1\rangle}}
\def\braket#1#2{\mathopen{\langle#1\,}|\mathclose {\,#2\rangle}}
\def\dspl{\displaystyle}
\maketitle
\begin{abstract}
A collection of the physical observables, related to the
electromagnetic properties of a nucleon, to investigate the
non--perturbative quantum fluctuations in the strong interaction
vacuum state under the influence of at least one close by (in energy
density) color superconducting phase found in several QCD motivated
model calculations, are studied. It is shown that the spontaneous
breaking of the electromagnetic gauge symmetry in the color
superconducting phase of strong interaction can result in relatively
clean signals in high energy processes, especially in the
semi-leptonic deep inelastic scattering ones, due to a kind of
electromagnetic induced strong interaction. A new type of mechanism,
which is a generalization of the Higgs one, through which the local
electromagnetic gauge symmetry is spontaneously broken by a
spontaneous breaking of the global baryon (nucleon) number
conservation, is revealed. A model independent assessment of the
question of how far is the color superconducting phase of the strong
interaction from its vacuum phase is made by studying currently
available experimental data on the electromagnetic responses of a
nucleon at high energies. It is shown that based on our current
knowledge about a nucleon, it is quite likely that there is at least
one color superconducting phase for the strong interaction that is
close enough to the vacuum state so that its effects can even be seen
in high energy processes besides heavy ion collisions.
\end{abstract}

\section{Introduction}

     Diquark condensation in the strong interaction ground states at
asymptotically high baryon density is a very like possibility
\cite{Bailin} since the dominant one gluon exchange interaction
between quarks in such a situation is attractive in the color triplet
channel which causes the BCS instability \cite{Son,Hsu} at the Fermi
surface for the quarks. At zero and low densities compared to the
nuclear saturation one, the study of the phenomenon in QCD, now called
{\em color superconductivity}, is more difficult due to the fact that
one has to deal with non-perturbative and finite density effects.  It
is quite apparent that the strong interaction vacuum is not color
superconducting at the present day condition in a large and uniform
region in space since color is known to be confined. The following
question can nevertheless be asked: {\em how far away (in energy
density) is certain type of metastable color superconducting phase,
called a virtual phase, of the hadronic vacuum from the stable one at
the present day condition?}.  It is speculated in
Refs. \cite{cspc2,cspc1,scalar-spc} that certain kind of color
superconducting phase exists, which has an energy density close to
that of the normal phase for the strong interaction vacuum in which
the chiral symmetry is spontaneously broken down.  Whether of not such
a speculation reflects reality is not a question that can be easily
answered theoretically based on QCD since unlike at high densities,
the QCD Lagrangian at low density is dominated by non-perturbative
effects. Lattice study of QCD at finite density is still facing
difficulties.  The properties of color superconducting phases inside a
nuclear matter with low or intermediate baryonic density was
investigated in the literature based on four fermion interaction
models \cite{cspc1,cspc2,scalar-spc,cspc3,cspc4}, which start 
in the early 90s and on
instanton motivated models \cite{Diak,Wilczek,Shuryak} recently, which is
currently actively studied.  However model studies will not be able to
determine with confidence when does certain kind of color
superconductivity will appear as the density increases.

The question can be answered using experimental means at high energy
due to the fact that for high energy processes, which probe the small
distance properties of the hadronic system, even a close by metastable
vacuum phase can contribute due to non-perturbative quantum
fluctuations that sample the contributions from the quasiparticles of
the metastable phase of the vacuum \cite{cspc3,FSpap,fdthry}.  Direct
detection of the virtual phase of the vacuum on the vacuum state
itself is very difficult if possible at all. Certain finite hadronic
system is needed to serve as a medium in order for significant effects
of the possible color superconducting virtual phase of the strong
interaction vacuum state to manifest. One of the best hadronic systems
for that purpose is a single nucleon. Besides the theoretical
simplicity of a nucleon, this is also mainly due to the fact that a
nucleon much better known experimentally compared to a heavy nucleus
so that anormalies with less uncertainties can be found.  On the
contrary, our combined theoretical and experimental knowledge about a
heavy nucleus a lot fuzzy for our purpose.

The effects of the possible virtual color superconducting phase of the
hadronic vacuum state can be observed as long as the virtual phase is
close enough to the stable normal phase for hadronic systems at zero
density and temperature. This is due to the fact that the
concentration of matter/energy inside of a nucleon generate
significant signal which is going to be discussed in detailed in the
sequel.  A nucleon in either of the above mentioned situations is
called a ``superconducting nucleon'' in the rest of the paper.  It
should not be confused with the naive pictures, which require that
there is in fact a true color superconducting phase or there are
certain kind of bound state of diquark or quark--quark clustering
inside of a nucleon, which, despite of the fact that are frequently
used, may be disfavored theoretically \cite{Gloz}.  Albeit this later
extreme pictures are logically covered by this paper they are not
neccessary ones (see also \cite{QNP2k}).  Work in the above mentioned
direction had been done in Ref. \cite{PCAC,PCAC1,GDHpap}, which lead
us to belief that there could be in fact a close by virtual color
superconducting phase for the strong interaction vacuum. This work
provides a more systematic and model independent study of the
possibility.

Since color is confined in nature at the present
day condition, it is extremely difficult, if possible at all, to
measure the color conductivity induced phenomenon of the hadronic
systems to make a direct assessment concerning whether or not a color
superconducting phase is contributing or not in a given hadronic
system. On the other hand, the electromagnetic (EM)
superconductivity
of a system can be probed using electrically charged particles
like charged leptons and hadrons since the electromagnetic charge is
not confined.

 An earlier study were carried out \cite{GDHpap} concerning the high
energy EM properties of a nucleon if it is superconducting in the
above mentioned sense.  The investigation is continued into the deep
inelastic scattering (DIS) processes of leptons and nucleons here by
trying to explain, in an essentially model independent way, a
collection of old and new anomalies about the nucleon in a logically
coherent way. These anomalies are about the peculiar behavior of the
elusive ``Pomeron'' in hadron--hadron collision and in DIS at small
Bjorken scaling variable $x$ and in certain exclusive processes. The
purpose of this work is not to contradict the current explanations of
some these problems in perturbative QCD. It tries to provide a
complementary view, based on Regge theory for strong interaction,
about the physical processes in the intersection area of soft and hard
scattering kinematic region.  Since this work, which contains no
detailed approximate computational scheme and model information, only
provides constraints due to symmetry considerations that any work
which properly implements the symmetries considered is expected to
arrive at basically the same results.

The small $x$ region of the nucleon structure function, where the
perturbative QCD calculation is no longer strictly valid, is an
intersection region in which both the perturbative and the
non-perturbative phenomena play their role. We do not understand the
non-perturbative QCD well enough from a first principle point of view
due to the lack of an effective and practical scheme to tackle the
problem. We do have, however, a phenomenologically successful one
based on Regge asymptotics, which determines the power law behavior of
the hadronic scattering amplitude in the large energy limits that
tells us something about the small and large $x$ behavior of the
structure function. Due to the constraint imposed by unitarity, the
physical hadronic amplitudes are bounded from above by the Froissart
bound which requires that the total hadron--hadron scattering cross
section $\sigma_{tot} \le \mbox{const}\times \log^2(s)$ with $s$ the
total energy. The Reggeon exchanged that is responsible for a behavior
like $\sigma_{tot}\sim \mbox{const}\times \log^2(s)$ is called the
soft Pomeron with an effective intercept of $\alpha_{\cal P}\approx
1.0$.  A high energy virtual photon interacts with the nucleon through
a coupling to the charged quarks inside the nucleon. Since quarks are
the basic building block of the nucleon, it is quite natural to assume
that the virtual photon and nucleon Compton scattering amplitude
respect the hadronic Froissart bound too, provided that the possible
final state phase space of these two reactions is of the same nature.

The recent experimental observation in lepton deep inelastic
scattering (DIS) and Drell--Yan processes involving a nucleon made it
possible to extract the unpolarized as well as polarized structure
functions $F_2$ \cite{H1,ZEUS,E665} and $g_1$ \cite{G1P,SMC,G1N}
respectively at small $x$ with large momentum transfer $Q^2$, the
flavor unsymmetry of the nucleon sea quark distribution \cite{E866}
and the charge symmetry breaking strength \cite{ChargeS,ChargeS2}.
The observed rapid rise of the structure functions in the small $x$
region was unexpected\footnote{More precisely, despite the fact that
it can somehow be accommodated by the perturbative QCD after the fact,
it was not predicted by it.} from a straight forward extrapolation of
the perturbative QCD pictures \cite{Eboli,LangCrit}. It turns out that
the soft pomeron can not provide the rapid rise of $F_2$ at small $x$
\cite{Pomeron1} either. Instead, for a $\gamma^* N$ scattering, there
is a transition region in the photon energy of order $1GeV$ after
which the so called hard pomeron with an intercept of $\alpha'_{\cal
P}\approx 1.4$ seems to be needed in order the reproduce the data
\cite{Abram}. Such a behavior can not continue all the way down to
$x=0$ if the hard pomeron is of hadronic nature since it would lead to
a violation of Froissart bound due to unitarity. Therefore although
two gluon model of the hard pomeron based on the QCD evolutoin
equations is capable of giving the proper rise of gluon density at
small $x$ \cite{Askow}, the above mentioned unitarity problem remains
in such an approach.  Although it can be argued that the true strong
interaction asymptotics is still far from our current accessible
energy scale, the physics dictated by the QCD BFKL evolution
equation may not be the whole story for the small $x$ physics for
reasons not related to its apparent violation of the Froissard bound
\cite{Eboli,LangCrit,Crit2}.

The observation that the center line of the flavor asymmetry can not
fully account for the violation of the Gottfried sum rule \cite{E866}
means that there seems to be less than three valence quarks inside a
nucleon \cite{Bertsch} when interpretated in a straight forward way.
Where is the missing quark number?  The charge symmetry breaking in
the nucleon structure function obtained in
Refs. \cite{ChargeS,ChargeS2} also needs to be explained.  The nucleon
structure function $F_2$ extracted from high precision neutrino DIS on
a nucleon \cite{Leung} disagree with the one extracted from the
charged lepton DIS on the same nucleon at small $x$ ($x < 0.1$). It is
still not clear what is the origin of this difference, which could be
caused by experimental systematic errors or it could be of physical in
nature as anticipated in Ref. \cite{PCAC}.  If the later possibility
is true, then how can such a behavior be understood?  In a recent
analysis of experimental data, the vector current conservation is in
question \cite{Close}. Can such a situation be incorporated into the
normal picture of hadronic system without certain drastic changes to
our faith to symmetries? Why does the polarized structure function
seems to change so rapidly which is reported in Ref. \cite{G1N} and
implied in Ref. \cite{SMC} at small $x$ that no theory \cite{G2Kolya}
seems can explain its origin at the present?

Admittedly, any individual piece of experimental information available
at the present may not be sufficiently accurate to lead to a final
conclusion, a coherent theoretical study of a collection of them is
expected to provide at least a stronger motivation for us to deepen
our understanding of the problem both theoretically and
experimentally.

 The physical processes behind the spontaneous partial breaking of the
EM local $U(1)$ gauge symmetry is discussed in section
\ref{sec:PBLGS}.  The manifestation of the possible spontaneous
partial breaking of the EM local gauge symmetry at not so high density
in the high energy semi-leptonic processes of hadronic systems is
studied in section \ref{sec:SLP}. The small $q^2$ region of the high
energy semi-leptonic scattering is discussed in section
\ref{sec:Q20}. The DIS region is discussed in section \ref{sec:DIS}.
An explanation of some of the current puzzles concerning the nucleon
in the DIS processes is provided.  Section \ref{sec:summary} contains
a discussion and a summary.

\section{Partial Breaking of A Local Gauge Symmetry}
\label{sec:PBLGS}

  The elucidation of the physical processes behind the spontaneous
breaking of fundamental local gauge symmetries provides an important
step for not only the construction of fundamental theories of nature,
like the successful standard model of electroweak interaction, but
also the further understanding of the physical properties of the
superconducting phase of certain condensed matter systems. The
historical development is covered in Ref. \cite{DGBbk}. The
discussion of the spontaneous breaking of the EM gauge symmetry and
that of the color $SU(3)$ gauge symmetry in the color superconducting
phase along the traditional lines are given, e.g., in
Refs. \cite{cspc2,cspc1,Raja,Cart}.  However the current understanding of
such a process, as it is briefly described in the following
subsection, stops at the situation in which the whole charge for the
local symmetry involved are spontaneously broken. With the deepening
of our knowledge about the structure of matter, especially the
hadronic systems, new situations emerge in which the above mentioned
mechanism becomes no longer suitable. They arise because the charges
for the corresponding fundamental local symmetries can be decomposed
as a superposition of various components that generate certain global
symmetries of the system. A familiar example is the electric charge of
a fundamental particle like a quark or a lepton.  It can be written
generically as
\begin{equation}
   \widehat Q_{em} = \widehat Q_B + \widehat Q_L + \widehat Q_V + \ldots
\label{Charge-decomp}
\end{equation}
in standard model, 
where $\widehat Q_B$ is the baryon number contribution to the total
electric charge, $\widehat Q_L$ is the lepton number contribution and
$\widehat Q_V$ is the isospin charge contribution, etc.

It is possible that the ground or vacuum state of the system does not
break the whole charge $Q_{em}$ but only its certain component which
generates global rather than local symmetry transformations, like part
of the baryonic charge $\widehat Q_B$ of the system.  In such a case
\cite{cspc1,cspc2,cspc3,scalar-spc}, the electromagnetic gauge
symmetry is said to be {\em spontaneously partial broken}.

But before embarking on understanding the physical processes behind
the spontaneous partial breaking of a local gauge symmetry, let us
briefly review the relatively well known Higgs mechanism for the
spontaneous full breaking of a gauge symmetry.

\subsection{A brief review of one charge local gauge theory and associated 
       physical states}
\label{sec:one-charge-rev}

The gauge transformation in a system possessing the corresponding
gauge symmetry can be classified into two categories 1) the gauge
transformation of the first kind and 2) that of the second
kind. Consider the simplest case of $U(1)$ locally symmetric gauge
theories. The symmetry transformations of the first kind correspond to
global ones generated by the charge operator
\begin{equation}
   \widehat Q = \int d^3x \widehat \rho(\mbox{\boldmath{x}},t=\mbox{const})
\label{U1Q1}
\end{equation}
of the system; the transformations of the second kind corresponding to
local realization of the global ones are generated by the Gauss
operator
\begin{equation}
  \widehat q(x) = \widehat \rho(x) - \nabla\cdot\widehat 
           {\mbox{\boldmath{E}}}(x),
\label{U1Q2}
\end{equation}
where $\widehat \rho(x)$ is the charge density operator and $\widehat
E(x)$ is the ``electric'' field operator. The existence of a local
gauge symmetry of the theory is characterized by the existence of a
superselection sector in the Hilbert space of the system, which is
called the space of physical states ${\cal H}_{phys} =\{ \ket{phys}
\}$, within which the matrix elements of $\widehat q(x)$ vanish,
namely
\begin{equation}
    \bra{phys}\widehat q(x) \ket{phys} = 0.
\label{phys-state1}
\end{equation}
Due to the local gauge symmetry of the system, the superselection
sector ${\cal H}_{phys}$ remains invariant during the time evolution
of the system----no transition from and to the other sectors of the
Hilbert space ${\cal H}'$, which are called the subspace of unphysical
states, is present.

In case of the gauge symmetry of the system is spontaneously broken
down, the vacuum state is not annihilated by the charge operator
$\widehat Q$, namely
\begin{equation}
   \widehat Q \ket{vac} \ne 0
\label{vacCtran}
\end{equation}
leading to massless Goldstone boson excitation of the symmetry
breaking according to the Goldstone theorem. These Goldstone bosons do
not belong to the physical superselection sector defined by
Eq. \ref{phys-state1} of the corresponding local gauge theory so that
the otherwise long range force generated by the Goldstone bosons
corresponding to a spontaneous breaking of the Global symmetry of the
first kind are actually absent in physical processes due to the
existence of the corresponding symmetry of the second kind generated
by Eq. \ref{U1Q2}. From the definition of the physical states, it can
be shown that the Goldstone bosons, generated by acting upon the
vacuum state by the total charge operator, belong to the subspace of
unphysical states ${\cal H}'$. Therefore the massless Goldstone bosons
associated with the spontaneous breaking of the symmetry
transformations of the first kind are absent in the physical processes
due to the gauge symmetry. The process of decoupling of the world-be
Goldstone bosons is often called the Higgs mechanism in high energy
physics.

\begin{figure}[h]
\begin{center}
\epsfbox{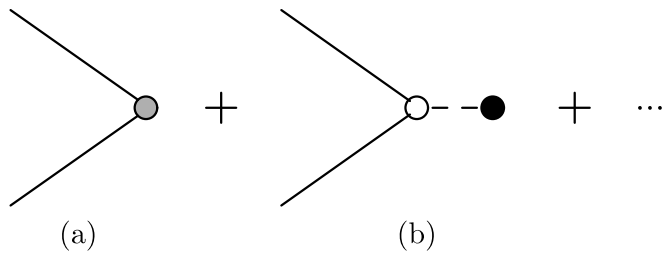}
\end{center}
\caption{\label{Fig:charge-spread}
  Various pieces for the charge of a fermionic (quasi-)particle. (a)
represents the core charge contribution that concentrates on the particle. The
strength of the core charge is smaller than the strength of the total charge of
the original particle. The piece of charge represented by (b) is spreaded. The 
dashed line in (b) denotes the propagator of the Goldstone boson.
}
\end{figure}

\begin{figure}[h]
\begin{center}
\epsfbox{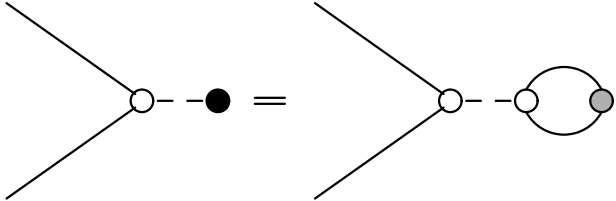}
\end{center}
\caption{\label{Fig:Jspread}
  The one loop graphical decomposition of the spreaded component of the 
  charge current in terms of elementary coupling constants. The hollow dots
  represents the Goldstone boson and fermion coupling vertices and
  the grey dot is the core part of the EM charge current vertex.}
\end{figure}

This mechanism can be presented in a way different from the more
familiar text-book discussion of the Higgs mechanism, which is based
upon a Lagrangian with Higgs fields appearing at the tree-level. We are
interested not only in such a possibility, but also the possibility
that the ``Higgs'' particles are not elementary ones but composite
excitations of the system that are absent at the tree-level. For that
purpose, a vertex functional representation is more appropriate. As it
is known, if a symmetry of the first kind is
spontaneously broken, then, the corresponding Goldstone bosons appear in
the vector current vertex of fermions, which is required by the
Ward--Takahashi identity relating the divergence of the vector current
to the self-energy of the fermions in the system \cite{cspc1}. In
this case, the charge of the fermion is only partially concentrated on
the fermionic (quasi-)particle, which forms a charge core, the rest of the
charge is spreaded around it due to the existence of the massless
world-be Goldstone boson. This is shown graphically in
Fig. \ref{Fig:charge-spread} and expressed in the following in terms
of current operator
\begin{equation}
    j^\mu(p+q,p) = j^\mu_{core}(p+q,p) + {q^\mu\over q^2}
    j_{sprd}(p+q,p) + \ldots,
\label{charge-sprd}
\end{equation}
where the core current is denoted as $j^\mu_{core}$ and the spreaded
piece of it, which has a strength characterized by a scalar function
$j_{sprd}$, is longitudinal with a massless pole in $q^2$. The
relative strength between the vector current $j_{core}^\mu$ and the
scalar one $j_{sprd}$ is determined by the current conservation,
namely,
\begin{equation}
    j_{sprd}(p+q,p) = -q_\mu j^\mu_{core}(p+q,p).
\label{jsprd-jcore}
\end{equation}
The Ward--Takahashi identity, which ensures the conservation of
charge, relates the self-energy and $j_{sprd}$ \cite{cspc1}.  It has
the form
\begin{equation}
   j_{sprd} = {e\over 2} \left [ O_3,\Sigma \right ] = \left (
         \begin{array}{cc} 0 & eD \\ -eD & 0 \end{array} \right )
\label{jsprd-Sig}
\end{equation}
in the 8-component ``real" representation for fermions with $e$ the
basic charge unit, $\Sigma$ the self-energy of the fermions, $O_3$ the
third Pauli matrix and the right hand side of the above equation
acting on the upper and lower four components of the 8-components
fermion spinor. The quantity $D$ is related to the order parameter for
the symmetry breaking. For a more concrete discussion, let us consider
the case of scalar fermion pair condensation studied in
Refs. \cite{scalar-spc,cspc3,Wilczek,Shuryak}. Define the Goldstone
boson--fermion coupling vertex as $ig\Gamma_S$ with $\Gamma_S$ given
by
\begin{equation}
  j_{sprd} = e\chi\Gamma_S
\label{GammaS}
\end{equation}
and $\chi$ the order parameter (see
Refs. \cite{scalar-spc,cspc3}). Then using the graphical representation
of the spreaded component of the current given in
Fig. \ref{Fig:Jspread}, one finds the value of the  Goldstone
boson--fermion coupling constant defined above to be
\begin{equation}
    g = \sqrt{\chi\over A}
\label{gvalue}
\end{equation}
with $A$ defined in the following loop diagram \cite{cspc1}
\begin{equation}
     -ieg^2 {q^\mu \over q^2} A \Gamma_S = \epsfbox{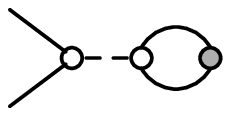} 
\end{equation}

\begin{figure}[h]
\begin{center}
\epsfbox{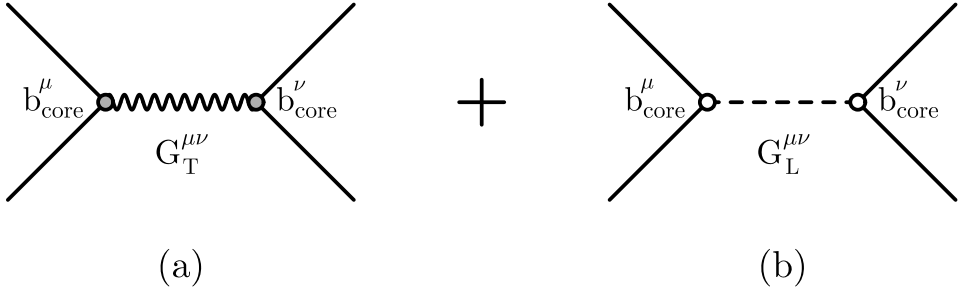}
\end{center}
\caption{\label{Fig:FFscattering}
The fermion--fermion scattering through vector coupling in the phase 
where the gauge symmetry of EM interaction is spontaneously broken down. Figure
(a) represents the scattering between particles with the same charge that
generates the spontaneous broken $U(1)$ symmetry by exchange a ``fully dressed
photon''. Figure (b) represents the scattering between
particles of the same type as figure (a) by the exchange of a 
Goldstone boson. The effective particle that mediates the
scattering of particles of the same type as figure (a) is obtained by
summing figures (a) and (b). The propagator of this effective 
combined excitation is a massive vector one.
}
\end{figure}

Since the photon couples to the vector current of the fermions, the
massless pole in the vector current vertex of the fermions in the
symmetry breaking phase of the system modifies the photon behavior
drastically. The fully dressed photon propagator $G_T^{\mu\nu}$
includes the self-energy insertion, namely,
\begin{eqnarray}
    G_T^{\mu\nu} &=& G_{0T}^{\mu\nu} + \left ( G_{0T}\Pi G_{0T} \right 
                     )^{\mu\nu} \\
   \Pi^{\mu\nu} &=& \pi^{\mu\nu}+\left ( \pi G_0 \pi \right )^{\mu\nu}
                      + \ldots \nonumber \\ &=&
          \epsfbox{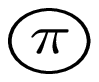} + \epsfbox{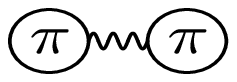} + \ldots
\label{Photo-SelfE}
\end{eqnarray}
in a diagrammatic representation with $\pi^{\mu\nu}=(g^{\mu\nu} q^2 -
q^\mu q^\nu )\pi$ the proper self-energy of the photon and $G_0^{\mu\nu}$
the photon propagator in free space in the gauge such that 
$q_\mu G_0^{\mu\nu}=0$.  In the phase
of the vacuum in which the EM $U(1)$ gauge symmetry is spontaneously
broken, the scalar quantity $\pi$ acquires a single pole in $q^2$,
namely, $\pi = m_\gamma^2/q^2 + \ldots $ with
\begin{equation}
    m_\gamma^2 = 4\pi \alpha_{em} g^2 A^2 = 4\pi \alpha_{em} \chi A
\label{photo-mass}
\end{equation}
depending on the order parameter of the symmetry breaking phase. Here
$\alpha_{em}$ is the fine structure constant. The Dyson equation for
the full photon propagator can be easily solved in the Lorentz
gauge. Keeping only the pole term of $\pi$, the photon propagator has
the following generic form in an infinite system
\cite{GDHpap,cspc1}
\begin{equation}
  G_T^{\mu\nu} = \left ( g^{\mu\nu} - {q^\mu q^\nu\over q^2} \right )
  {-i \over q^2 - m_\gamma^2 + i \epsilon}.
\label{Photon-prop}
\end{equation}
The fermion--fermion scattering in the vector channel represented by
Fig. \ref{Fig:FFscattering} is mediated by the exchange of the media
modified photon with its propagator given by
Eq. \ref{Photon-prop}. From Eqs. \ref{charge-sprd} and
\ref{Photon-prop}, it follows that only the core part of the current 
density $j_{core}^\mu$ participates in the scattering. Namely, the
scattering amplitude $T_{fi}$ is given by
\begin{equation}
   -i T_{fi}^{(T)} =  (ij_{core})_\mu G_T^{\mu\nu} (ij'_{core})_\nu,
\end{equation}
which is generated by ``less'' charge (compared to that of the
original free fermions). Therefore the scattering strength between
fermions is modified by the symmetry breaking.

The spread charge current density $j_{sprd}$ also contribute to the
fermion--fermion scattering; it is mediated by an exchange of the
massless  Goldstone boson. Using the $U(1)$ Ward--Takahashi
identity for EM \cite{cspc1}, it can be shown that the 
Goldstone boson contribution to the fermion--fermion scattering is of
the form
\begin{equation}
   -i T^{(L)}_{fi} = \bra{f} (ig\Gamma_S)\ket{i} {i\over q^2}\bra{f'}
              (ig\Gamma_S)\ket{i'} = 
              (ij_{core})_\mu G_L^{\mu\nu} (ij'_{core})_\nu
\label{TLdef}
\end{equation}
with
\begin{equation}
    G_L^{\mu\nu} = {iq^\mu q^\nu \over m_\gamma^2 (q^2 + i\epsilon)},
\label{GLmn}
\end{equation}
which is represented in Fig. \ref{Fig:FFscattering}.b. Here
Eqs. \ref{GammaS}, \ref{jsprd-jcore}, \ref{gvalue} and
\ref{photo-mass} are used. Therefore, the total scattering amplitude
is $T_{fi} = T^{(T)}_{fi} + T^{(L)}_{fi}$, which is of the form
\begin{equation}
    -iT_{fi} = (ij_{core})_\mu G^{\mu\nu} (ij'_{core})_\nu
\end{equation}
with the full propagator $G^{\mu\nu}$ given by
\begin{equation}
     G^{\mu\nu} = G_T^{\mu\nu} + G_L^{\mu\nu} = \left (g^{\mu\nu} - {q^\mu q^\nu\over m_\gamma^2} \right )
                         {-i\over q^2 - m_\gamma^2 + i\epsilon}.
\label{Pfull-prop}
\end{equation}
The effective full propagator $G^{\mu\nu}$ for the interaction is the
standard one for a massive vector particle with mass $m_\gamma^2$.

Before proceeding to the next subsection, let us make a few remarks:
1) although the core current density is used in the scattering
amplitude $T_{fi}$, which weakens the interaction vertex, the particle
exchanged has an additional longitudinal component. It is in this way
that the gauge symmetry is maintained 2) the massless pole
corresponding to the Goldstone boson and the massless pole of the
photon disappear in the final physical scattering
amplitudes. Therefore the originally non-existent Goldstone boson in
the Lagrangian do not appear in the final results either; it only
appear in the intermediate steps of the discussion. So it does not
belong to the subspace physical states. What exchanged in the
fermion--fermion scattering is a massive vector boson with standard
propagator Eq. \ref{Pfull-prop} for {\em any charged particles inside
of the system} 3) it is evident from Eq. \ref{TLdef} that the coupling
constant $g$ between the Goldstone boson is combined with the original
massless photon to generate an effective massive vector
excitation. Despite that the Goldstone boson couples to the fermions
with a strength $g$ which is of order 1 due to the fact that the
spontaneous partial breaking of the EM gauge symmetry is caused by
strong interaction, this massive vector particle only couples to the
fermions with a strength of EM interaction due to a delicate
cancellation of its strong interaction effects.

\subsection{Partial breaking of local gauge symmetry in multiple charge 
systems}

The charge operator in Eq. \ref{U1Q1} contains only one piece since
the system understudy contains only one charge
\footnote{For example, the gas of electrons in condensed matter
systems and the spontaneous symmetry breaking process in the standard
model of the electroweak interaction. The spontaneous breaking of the
color $SU(3)$ gauge symmetry in the color superconducting phase of
quarks system interested in this work also belong to this
category!}. The electric charge operator in the realm of elementary
particle physics contains multiple pieces. In the standard model, the
electric charge operator can be decomposed like in Eq.
\ref{Charge-decomp}. In addition, the baryon number contribution
$\widehat Q_B$ can be further decomposed into components associated
with each of the three generations of hadrons, namely
\begin{equation}
    \widehat Q_B = \widehat Q_N + \ldots
\label{QBdecomp}
\end{equation}
with $\widehat Q_N$ the contributions of the first generation of
quarks consists of the light up (u) and down (d) quarks to the baryon
number. The electric charge density operator can be decomposed in the
same way. Although the symmetry generated by $\widehat Q_{em}$ has an
associated local gauged one with a superselection sector of physical
states, its components do not always has a local symmetry attached to
them.

If there is a spontaneous breaking down of a (global) symmetry
corresponding to any one component of the electric charge, then the
associated Goldstone bosons do not necessarily belongs to the subspace
of unphysical states (for the EM gauge symmetry) and they are expected
to participate in the physical scattering processes in certain
reaction channels. The Goldstone bosons still can has certain overlap
with the physical superselection sector for the EM gauge symmetry in
the Hilbert space of the system after the unphysical component is
projected out using Eq.\ref{phys-state1}.

  As it is discussed above, the Goldstone boson for a single charge
system is not a physical excitation. The effective excitation that
mediates the interaction between the charged particles is a massive
vector boson. The massless excitations decouple from the physical
spectra. These occur due to a delicate cancellation of the long range
interaction effects between the longitudinal components of
$G_T^{\mu\nu}$ and $G_L^{\mu\nu}$ that couples to the emitter and
receiver with a strength in the range of the {\em strong interaction}
rather than that of the EM interaction. Such a cancellation is
protected by the gauge invariance against higher order corrections.

  For a multi-charged system like the standard model, the situation
becomes a little complicated. For the scattering between two particles
with the same broken charge (e.g., the nucleon charge in the light
quark system), the cancellation between the long range interaction
between $G_T^{\mu\nu}$ and $G_L^{\mu\nu}$ still occur
\cite{cspc1}. Therefore a particle experiences a short ranged EM
interaction when meets a particle with the same broken charge in the
superconducting phase. In this case, the Higgs mechanism is
functioning.  What happens if two particles, A and B, with different
broken charge meet?  Assuming that particle A radiates a photon with a
strength determined by its own electric charge and a Goldstone boson with
a strength determined by its broken charge\footnote{It is ultimately
determined by the pattern of symmetry breaking specified by the mass
matrix, see Eq. \ref{jsprd-Sig}}.  The photon will propagate with
medium modified propagator $G_T^{\mu\nu}$ on its way to particle B and
the Goldstone boson will propagate with propagator $G_L^{\mu\nu}$ on
its way also to particle B. Particle B receives the photon with the
response determined by its electric charge and the Goldstone boson
with the response determined by its own broken
charge. Cancellation between the two forces induced by the longitudinal
component of the medium modified EM field and that of the Goldstone
boson does not happen. The residue force couples particle A and B with
a strength of that of strong interaction since the Goldstone boson 
couples A and B in such a way. The same phenomenon happens
if the role of A and B is exchanged.

This can be putted in another way. In a multi-charged system in its
superconducting phase which breaks the EM gauge symmetry, the {\em
effective} EM field strength produced by a charge particle is {\em
observer dependent}. For a particle with the same broken charge as the
source, it sees, effectively, a field of massive (short ranged in the
static limit) vector particle with an strength of EM interaction.  For
a particle with different broken charge, what it effectively sees is
two force fields: the first one is a field of massive vector particle
with a intensity of EM interaction and the second one is a residue
scalar field (with gradient coupling to charged particles) with a
intensity comparable to that of the strong interaction that is
responsible for the symmetry breaking.

Let us discuss the above statements in a more
concrete fashion using the language of the previous subsection. First
the Fock space of the system at the tree level is divided into
subspaces, each of which contains particles with the same charge, say
$\widehat Q_N$, that is spontaneously broken. Denoting the total EM
charge current density as $J^\mu = b^\mu + \overline J ^\mu$ with
$b^\mu$ the contribution of the charge that is spontaneously broken
and $\overline J ^\mu$ the remaining charges that are unbroken. In the
symmetry breaking phase, the corresponding vertex $b^\mu(p+q,p)$ for
the charge current density $b^\mu$ can be decomposed in the same form
as Eq. \ref{charge-sprd} so that the total vertex for the EM charge
current density $J_\mu(p+q,p)$ has a form
\begin{equation}
 J^\mu(p+q,p) = \left [ b_{core}^\mu(p+q,p)+{q^\mu\over
 q^2}b_{sprd}(p+q,p) \right ] + \overline J ^\mu(p+q,p)+\ldots.
\label{Jdecomp}
\end{equation}
Here, the vertex for the broken charge density operator $b^\mu$ is
decomposed into a sum of the core part and the spreaded part. The rest
of the vertices corresponding to the unbroken charge current density
is denoted by $\overline J ^\mu(p+q,p)$.  Then, using the method
of the previous subsection, it is not hard to show that the 
fermion--fermion scattering amplitude $T_{fi}$ in the lowest order
approximation in the fine structure constant has the following general
form
\begin{equation}
   -iT_{fi} = (ib_{core})_\mu G^{\mu\nu} (ib'_{core})_\nu +
                (ib_{core})_\mu G_T^{\mu\nu} (i\overline{J'})_\nu +
                (i\overline J)_\mu G_T^{\mu\nu} (ib'_{core})_\nu +
                (i\overline J)_\mu G_T^{\mu\nu} (i\overline {J'})_\nu,
\label{Tfi1}
\end{equation}
where $J_\mu$, $J'_\nu$ denote the matrix elements of the EM current
operator and $G_T^{\mu\nu}$ and $G^{\mu\nu}$ are given by
Eqs. \ref{Photon-prop} and \ref{Pfull-prop} respectively with
$m_\gamma^2\sim e_S^2$ where $e_S$ is the broken charge of the
charged particle. Eq. \ref{Tfi1} indicates that only the broken charge
components of the fermions are scattered by an exchange of an authentic
massive vector excitation, the scattering between the rest components
of the EM charge of fermions and between the broken component and the
unbroken components of the EM charge operator of the fermions are
mediated by exchange of massive vector excitation with its
longitudinal component $G_L^{\mu\nu}$ given by Eq. \ref{GLmn}
removed. This effective excitation contains a mixture of massive and
massless excitation, which can be read out directly from
Eq. \ref{Photon-prop}. The massless pole in $G_T^{\mu\nu}$ will not be
present explicitly on the right hand side of Eq. \ref{Tfi1}
in the vacuum case since 
$\overline J_\mu$ is a conserved current. This provides us the freedom to
modify the $q^\mu q^\nu$ term of $G_T^{\mu\nu}$. It is quite natural
to write Eq. \ref{Tfi1} in an equivalent form, namely,
\begin{eqnarray}
  -iT_{fi} &=& (ib_{core})_\mu G^{\mu\nu} (ib'_{core})_\nu +
                (ib_{core})_\mu G^{\mu\nu} (i\overline{J'})_\nu +
                (i\overline J)_\mu G^{\mu\nu} (ib'_{core})_\nu +
                (i\overline J)_\mu G^{\mu\nu} (i\overline {J'})_\nu,
\nonumber \\
  &=& (ib_{core} + i \overline J)_\mu G^{\mu\nu} (ib'_{core}+ i \overline J)_\nu
   = (iJ_{core})_\mu G^{\mu\nu} (iJ_{core})_\nu,
\label{Tfi2}
\end{eqnarray}
where the propagator $G_T^{\mu\nu}$ in Eq. \ref{Tfi1} is replaced by
the full propagator $G^{\mu\nu}$.

From Eq. \ref{Tfi2}, it can be seen that in the fermion--fermion EM
interaction inside the system, the core part of the EM charge
participate in the interaction between charged particles exchange a
massive vector boson. Since the core part of the broken charge current
component of the EM charge current is not conserved, the formally
similar four terms on the right hand side of Eq.  \ref{Tfi2} has
different physical meaning. In the first term, the longitudinal
component of $G^{\mu\nu}$ contributes since $b^\mu_{core}$ is not
conserved by itself. This term can be interpreted as that both the
core component and the spread component of the broken charge
participate in the fermion--fermion interaction in this particular
channel.  The second and third terms on the right hand side of
Eq. \ref{Tfi2} have quite different physical properties. This is
because $\overline J_\mu$ is a conserved quantity the longitudinal
component of $G^{\mu\nu}$ does not actually participate in the
fermion--fermion scattering in the vacuum. This is one of the reasons
why it is difficult to detect the effects of the Goldstone boson in
the relativistic invariant vacuum state of strong interaction.

In case of a nucleon or in a nuclear medium, the contraction of
$G_L^{\mu\nu}$ with $\overline J_\mu$ does not give vanishing result
(see the following discussion). So Eq. \ref{Tfi1} should be used.
The absence of the longitudinal contribution also means that the
effects of the massless Goldstone boson is observable. Therefore we do
not have a full Higgs phenomenon; the gauge symmetry is therefore
considered to be ``spontaneously partial broken''.  Let us consider the
lowest order (in $\alpha_{em}$) inclusive semi-leptonic cross section. 
The scattering cross section $\sigma_{in}$ is related to the imaginary
part of the corresponding forward Compton scattering amplitude
$iT^{\mu\nu}$,
\begin{equation}
  -iT^{\mu\nu} = Z_\gamma^{-2} q^4 \int d^4x e^{iq\cdot x}\bra{N}
   TA^\mu(x) A^\nu(0)\ket{N},
\end{equation}
where ``T'' denotes time ordering and $G^{\mu\nu}_{N;\gamma}=\bra{N}
TA^\mu(x) A^\nu(0)\ket{N}$ is the photon propagator in the presents of
a nucleon \cite{GDHpap}. The amputation of the external photon line
{\em outside} of the nucleon is represented by the factor
$Z_\gamma^{-2}q^4$ with $Z_\gamma$ the photon wave function
renormalization constant.  The discussion of subsection
\ref{sec:one-charge-rev} for the modification of the photon propagator
(see also Ref. \cite{GDHpap}) is also applicable for
$G_{N;\gamma}^{\mu\nu}$. But there are two major difference.

First, the vacuum state of the strong interaction is an infinite
system in spatial extension whereas a nucleon is a finite size
system. So only when the wave length of the photon is much smaller
than the size of a nucleon, the photon propagator would has a similar
behavior to the vacuum one \cite{GDHpap}. Otherwise, any pole behavior
in the propagators is smeared.  For the simplicity of the discussion, such
a smearing effects are not mentioned explicitly in most part of the
discussions except in the small $Q^2\equiv -q^2$ region since we are
discussing high energy processes in which the wave length of the
(virtual) photon is much smaller than the size of a nucleon.

Second, the vacuum state of the strong interaction is an isoscalar so
that the proper self-energy $\pi^{\mu\nu}$ for the vacuum state,
\begin{eqnarray}
    -i\pi^{\mu\nu} (q) &=& \left .\int d^4 x e^{iq\cdot x} \bra{0} T J^\mu(x)
                     J^\nu(0) \ket{0}\right |_{1PI} \nonumber \\ 
  &=& \left .\int d^4 x e^{iq\cdot x} \bra{0} T \left [b^\mu(x) b^\nu(0) +
                     \overline J^\mu(x) \overline J^\nu(0) \right ]
                     \ket{0}\right |_{1PI}
\end{eqnarray} 
contains no interference term between the current of broken charge and
the current of the unbroken charge due to the isospin symmetry of the
vacuum state. For the simplicity of the notation, we shall omit the
``1PI'' (one photon irreducible) subscript in the following.  Since
the proper self-energy $\pi_N^{\mu\nu}$ for a photon inside a nucleon
is induced by the interaction in an isospin unsymmetric background
provided by the nucleon \cite{E866}, the general form of the photon
proper self-energy inside a nucleon contains two additional terms
corresponding to the interference terms between the current of the
broken charge and that of the unbroken charge, namely,
\begin{eqnarray}
    -i\pi_N^{\mu\nu} (q) &=& \int d^4 x e^{iq\cdot x} \bra{N} T \left
              [b^\mu(x) b^\nu(0) + b^\mu(x)
              \overline J^\nu(0) 
              + \overline J^\mu(x)
              b^\nu(0) + \overline J^\mu(x) \overline J^\nu(0)
                                    \right ] \ket{N}.
\end{eqnarray} 

   The term for the broken charge of a nucleon is
\begin{eqnarray}
  \bra{N}T b^\mu(x) b^\nu(0) \ket{N}& =& \bra{N} \left [Tb^\mu_{core}(x)
                         b^\nu_{core}(0) + {\widehat
                         q^\nu\over\widehat q^2} Tb_{core}^\mu(x)
                         b_{sprd}(0) \right . \nonumber 
\\              
&&\hspace{0.6in} \left .
                   + {\widehat q^\mu\over\widehat
                         q^2} Tb(x)_{sprd} b^\nu_{core}(0) - {\widehat
                         q^\mu\widehat q^\nu \over\widehat q^4} T
                         b_{sprd}(x) b_{sprd}(0) \right ] \ket {N} +
                         \ldots ,
\label{Nbbdecomp}
\end{eqnarray}
where $\widehat q_\mu = i \partial_x^\mu$ acts
only on $b_{sprd}$ in the above equations. It can be seen that in
order to maintain gauge invariance, the correlator contains not only a
single massless Goldstone boson pole, but also a double massless pole
that is absent in the vacuum case.

The second and third interference terms have isospin odd component
that is present in a single nucleon state.  Similar to
Eq. \ref{Nbbdecomp}, they can be decomposed as
\begin{eqnarray}
     \bra{N} T b^\mu(x) \overline J^\nu(0) \ket {N} &=& \bra{N} 
                   \left [ Tb^\mu_{core}(x)\overline J^\nu(0) +
                   {\widehat q^\mu\over \widehat q^2} Tb_{sprd}(x)
                   \overline J^\nu(0) \right ] \ket{N} + \ldots,
\label{IsoOdd1} \\ 
     \bra{N} T \overline
                   J^\mu(x)b^\nu(0) \ket {N} &=& \bra{N}  \left [
                   T\overline J^\mu(x)b^\nu_{core}(0) + {\widehat
                   q^\nu\over \widehat q^2} T\overline J^\mu(x)
                   b_{sprd}(0) \right ] \ket{N} + \ldots .
\label{IsoOdd2}
\end{eqnarray}
Only a single massless  Goldstone pole is present for these
isospin odd terms.

There is an additional important feature compared to
the vacuum case due to the fact a nucleon behaves like a medium for the
photon. The matrix element of the spread current in the momentum space
should be of the following form
\begin{equation}
     b_{sprd}^\mu(p+q,p) = {q^\mu + \beta P^\mu\over q^2+\beta
                            q\cdot P} b_{sprd}(p+q,p)
\label{med-bsprd}
\end{equation}
instead of the simple form given by Eq. \ref{charge-sprd}. Here
$P^\mu$ is the 4-momentum of the nucleon $\ket{N}$ (the medium),
$p^\mu$ is the 4-momentum of a quark. The coefficient $\beta$ is
determined by the nature of the symmetry breaking inside the
nucleon. $\beta$ is expect to decrease at large $q\cdot P$ as
\begin{equation}
   \beta \sim {m^{-1}\mu^3\over {q\cdot P}}
\label{beta-decr}
\end{equation}
which can be estimated by a conventional finite density computation,
with $\mu$ certain mass scale characterizing the concentration of a quark
matter inside the nucleon and $m$ is the mass of the nucleon. There is
a corresponding change to the propagator $G_T^{\mu\nu}$ for the photon
in the nucleon, which is also characterized by $\beta$. It should not
be discussed quantitatively here.  What is important here is the
additional term $\beta P^\mu$ in the definition of the
``longitudinal'' Goldstone boson polarization, namely
\begin{equation}
l^\mu = q^\mu + \beta P^\mu. 
\label{longt}
\end{equation}
The medium term $\beta P^\mu$, together with the Goldstone boson,
are responsible for the ``extra'' particle production in a high energy
semi-leptonic process involving a nucleon.

In most of the expressions in the following, instead of $l^\mu/q\cdot
l$, $q^\mu/q^2$ is still used for the Goldstone boson contributions
for notational simplicity. But care should be taken that the
contraction of such a ``$q^\mu$'' which is actually $l^\mu$,       with a
conserved current should not be replaced by a zero.

\subsection{The general picture}
\label{sec:general-pic}

   Two aspects are important here.  The first one is related to the
phenomenon of missing charge. In certain channels of reaction, like
the semi-leptonic processes, only the core charge of the hadronic
system couples to the external probes. Since a finite percent of the
broken charge of particle is removed from its core component and added
to its spreaded component in the symmetry breaking phase of the
system, the system appears to these probes to contain less charge than
it would normally be expected \cite{PCAC}.

\begin{figure}[h]
\begin{center}
\epsfbox{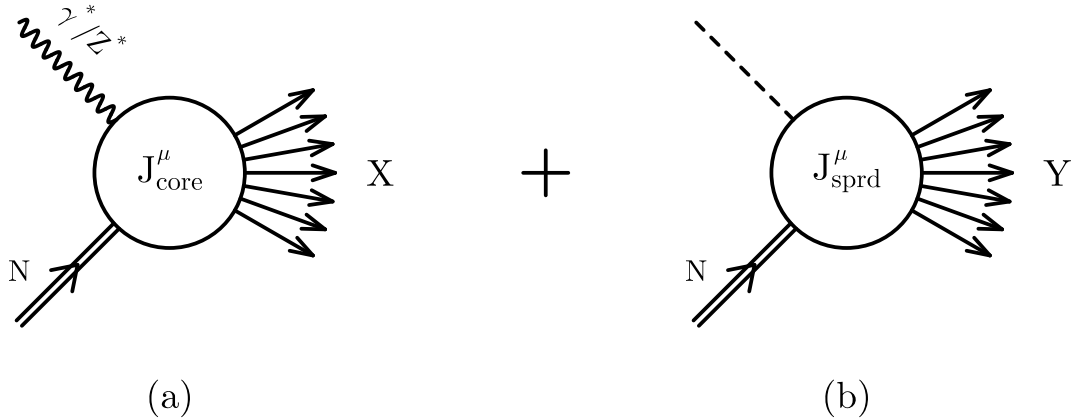}
\end{center}
\caption{\label{Fig:finals} The final states generated in the neutral
current semi-leptonic processes.  Figure (a) represents the final
states ``X'' generated by absorbing a (medium modified) photon with
propagator given by Eq. \ref{Photon-prop}. Figure (b) represents final
states ``Y'' generated by absorbing a Goldstone boson.  The strength
of both of these two couplings are in fact of order of the strong
interaction. If both of the photon and the Goldstone boson are emitted
by another nucleon, then proper cancellation between these two sets
diagrams occurs. The result is an effective massive vector photon that
couples to the nucleon with a strength of EM interaction which
generates far less final states compared to all other cases in which
this kind of cancellation does not happen. }
\end{figure}

The second one is connected to the  Goldstone boson degree of
freedom. Let us imagine that a superconducting nucleon is hit by a
high energy virtual photon $\gamma^*$ or $Z$ boson like the one
shown in Fig. \ref{Fig:finals}.a, then the quarks (quasi-particles)
start to radiate gluons, photons and  Goldstone
bosons. Because $G^{\mu\nu}=G_T^{\mu\nu}+G_L^{\mu\nu}$ couples to the
quarks inside the nucleon with a strength of EM interaction and
$G_L^{\mu\nu}$ couples to the quarks with a strength of the strong
interaction which is two order of magnitude larger that the EM one,
$G^{\mu\nu}_T$ must also contain an corresponding component that
couples to the quarks with a strength of the strong interaction in the
superconducting phase. Therefore, the total cross section generated in
Fig. \ref{Fig:finals}.a is larger then that of a typical EM
interaction in normal situations. For the same reason, the total cross
section generated in Fig. \ref{Fig:finals}.b is also larger then that
of a typical EM interaction. 

These two sets of graphs cancel each other if both of them are emitted
by a nucleon which also contains up and down quarks. In this case, the
Higgs mechanism discussed in subsection \ref{sec:one-charge-rev} is
realized. The nucleon--nucleon (NN) interaction is normal strong
interaction plus the normal EM interaction mediated by a massive
vector photon. Since the strength of the EM interaction is much
smaller than the strong interaction strength, the total cross section
produced by the summation of Figs. \ref{Fig:finals}.a and
\ref{Fig:finals}.b is much smaller than that of other strong
interaction processes.

In the case of charged lepton and nucleon scattering, on the other
hand, the charged leptons does not couple to the  Goldstone boson.  
So only Fig. \ref{Fig:finals}.a is left. Let us write
\begin{equation}
  G_T^{\mu\nu} = G^{\mu\nu} - G_L^{\mu\nu}
\end{equation}
with $G^{\mu\nu}$ the massive photon propagator that couples to 
the quarks with a strength of EM interaction. One can ignore
$G^{\mu\nu}$ in this case and write
\begin{equation}
  G_T^{\mu\nu} \approx - G_L^{\mu\nu}.
\label{GTstrng}
\end{equation}
In the vacuum, since $G_L^{\mu\nu}$ is only proportional to $q^\mu
q^\nu$, $G_L^{\mu\nu}$ can not mediate force between charged leptons
and quarks due to the current conservation on the lepton vertex.  But
inside a nucleon, $G_L^{\mu\nu}$ is proportional to $l^\mu l^\nu$. It
contains terms like $\beta (P^\mu q^\nu + P^\nu q^\mu)$ and $\beta^2
P^\mu P^\nu$ due to Eq. \ref{med-bsprd}. These terms do not produce a
vanishing result when contracted with the lepton current. In addition,
the contraction of the nucleon 4-momentum $P^\mu$ with the lepton or
quark current generate a factor $q\cdot P$ at large photon
energy. Therefore, due to the medium effect, the strong interaction
force mediated by the Goldstone boson is independent of any
mass scale even in high energy reactions. Such a strong interaction
force does not, however, exist between nucleons.

Therefore, {\em there appears to be a new type of strong interaction
inside a nuclear system when probed by systems other than another
nuclear systems if a nucleon (or a nucleus) is superconducting}.  As
it is shown in the following, this new kind of strong interaction
force may be identified with the so called ``hard pomeron''.

This specific mechanism for the explanation of the violation of the
Froissart bound has several testable predictions which are explicated
in the sequel.

\section{Semi-Leptonic Processes at High Energies}
\label{sec:SLP}

\begin{figure}[h]
\begin{center}
\epsfbox{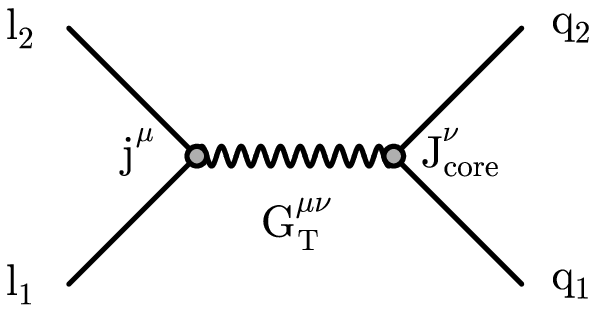}
\end{center}
\caption{\label{Fig:SemiLP} The lepton--quark scattering when the EM
gauge symmetry is spontaneous broken. In this case, the effective
vector excitation exchanged contains not only a massive component
$G^{\mu\nu}$, but also a massless component $-G_L^{\mu\nu}$.  The
contribution from the exchange of the Goldstone boson is absent.  What
the lepton sees is the core component of the broken quark baryon
charge of the light quarks.}
\end{figure}

One way of observing the partial breaking of the EM gauge symmetry
inside a hadronic system is to use the lepton--hadron
scattering. Since leptons do not couple to the corresponding Goldstone
boson. The general scattering amplitudes given by Eq.
\ref{Tfi1} is simplified
\begin{equation}
  -iT_{fi} = (ib_{core}+ i\overline J)_\mu G_T^{\mu\nu} (ij)_\nu =
   (iJ_{core})_\mu G_T^{\mu\nu} (ij)_\nu,
\label{Semi-lp-T}
\end{equation}
where $j_\mu$ is the electro--weak current of the leptons and
$J_{core}^\mu = b_{core}^\mu + \overline J^\mu$. So only the core part
of the electro--weak current of the hadrons contributes to the
semi-leptonic processes. The spreaded component of the broken charge
of the nucleon is invisible. This is diagrammatically represented by
Fig. \ref{Fig:SemiLP}.

The process of the inclusive semi-leptonic DIS process is connected to
the imaginary part of the forward virtual photon $\gamma^*$ and $Z^*$-
nucleon Compton scattering amplitude
\begin{equation}
   -i T_{fi}^{\mu\nu}(p,q) = \int d^4 x e^{iq\cdot x}\bra{N} T J^\mu(x) J^\nu(0) \ket{N}, 
\label{DISTmn}
\end{equation}
which can be decomposed according to Eq. \ref{Nbbdecomp}.

\subsection{$lN$ neutral current scattering processes}

\subsubsection{Gauge invariance,scattering amplitudes and structure functions}

   The EM interaction in the semi-leptonic $lN$ scattering conserves
parity. The kind of parity conserving decomposition of
Eq. \ref{DISTmn} is written as a sum of parity even invariant
amplitudes \cite{GDHpap}
\begin{eqnarray}
  T^{\mu\nu}(p,q) &=& {1\over 2m } \overline U(pS) \left [ H_1
    g^{\mu\nu} - H_2{1\over m^2} q^\mu q^\nu + H_3 {1\over m^2} p^\mu
    p^\nu - H_4{1\over m^2} (p^\mu q^\nu + p^\nu q^\mu ) +
    iH_5\sigma^{\mu\nu}
\right . \nonumber \\
&&\left .  
+ iH_6{1\over m^2}
(p^\mu\sigma^{\nu\alpha}q_\alpha-p^\nu\sigma^{\mu\alpha}q_\alpha) +
iH_7{1\over m^2}
(q^\mu\sigma^{\nu\alpha}q_\alpha-q^\nu\sigma^{\mu\alpha}q_\alpha) +
iH_8 {1\over m^3} \epsilon^{\mu\nu\alpha\beta}q_\alpha
p_\beta\rlap\slash q \gamma^5 \right ] U(pS),
\label{InvAmpDecomp1}
\end{eqnarray}
where $U(pS)$ is the spinor for a nucleon with momentum $p$ and spin
$S$ and $H_i$ ($i=1,2,\ldots,8$) are invariant amplitudes. If the EM
gauge symmetry is spontaneously partial broken, then some of these
amplitudes with at least one $q^\mu$ and/or $q^\nu$ in front carrying
the Lorentz indices $\mu$ or $\nu$ develop single and double
poles\footnote{They are $l^\mu$, and $l\cdot q$ respectively for a
nucleon in more precise sense. But we shall not make the distinction
between them since the difference can be removed by a proper linear
combination of $H_i$. So the generality of the discussion will not be
affected.} in $q^2$. These parity even invariant amplitudes are $H_2$,
$H_4$ and $H_7$. $H_2$ is the coefficient of $q^\mu q^\nu$, so apart
from others, it constraints the contributions from the fourth term in
Eq.  \ref{Nbbdecomp} which allows it to have a double pole in
$q^2$. $H_4$ and $H_7$ are coefficient of those terms that are linear
in $q^\mu$ or $q^\nu$, they contain contributions from the second and
third terms of Eq.  \ref{Nbbdecomp} and the second terms of
Eqs. \ref{IsoOdd1} and \ref{IsoOdd2}.  Therefore, the poles of these
three invariant amplitudes can be separated out
\begin{eqnarray}
   H_2 &=& \overline H_2 + {m\over 2 q^2} a - {m^2\nu\over q^4} b, \label{H2decomp}\\
   H_4 &=& \overline H_4 + {m\over q^2} b ,\label{H4decomp}\\
   H_7 &=& \overline H_7 + {m\nu\over q^2} c. \label{H7decomp}
\end{eqnarray}
The invariant amplitudes $H_i$ ($i=1,2,\ldots,8$) are not independent
of each other due to the gauge invariance expressed in terms of the
Ward identity
\begin{equation}
    q_\mu T^{\mu\nu} = 0,
\end{equation}
which imposes constraints on them, namely,
\begin{eqnarray}
   m^2 H_1(q^2,\nu) - q^2\overline H_2(q^2,\nu) - m\nu \overline
   H_4(q^2,\nu) &=&{1\over 2} m a(q^2,\nu),
\label{HBZ1} \\
   m\nu H_3(q^2,\nu) - q^2 \overline H_4(q^2,\nu) &=& mb(q^2,\nu),
\label{HBZ2}\\
    m^2 H_5(q^2,\nu) - m\nu H_6(q^2,\nu)-q^2\overline H_7(q^2,\nu) &=& m\nu c(q^2,\nu).
\label{HBZ3}
\end{eqnarray}
As discussed above that in the semi-leptonic processes, only the core
part of the scattering amplitude are observable. The core part of the
invariant amplitude $T^{\mu\nu}$ is obtained from it by removing the
poles in $q^2$. So
\begin{eqnarray}
  T_{core}^{\mu\nu}(p,q) &=& {1\over 2m} \overline U(pS) \left [ H_1
    g^{\mu\nu} - \overline H_2{1\over m^2} q^\mu q^\nu + H_3 {1\over
    m^2} p^\mu p^\nu - \overline H_4{1\over m^2} (p^\mu q^\nu + p^\nu
    q^\mu ) + iH_5\sigma^{\mu\nu}
\right . \nonumber \\
&&\left .  + iH_6{1\over m^2}
(p^\mu\sigma^{\nu\alpha}q_\alpha-p^\nu\sigma^{\mu\alpha}q_\alpha) +
i\overline H_7{1\over m^2}
(q^\mu\sigma^{\nu\alpha}q_\alpha-q^\nu\sigma^{\mu\alpha}q_\alpha) +
iH_8 {1\over m^3} \epsilon^{\mu\nu\alpha\beta}q_\alpha
p_\beta\rlap\slash q \gamma^5 \right ] U(pS),
\label{InvAmpDecomp2}
\end{eqnarray}
which, after considering the constraints Eqs. \ref{HBZ1}--\ref{HBZ3},
can be reduced to
\begin{eqnarray}
  T_{core}^{\mu\nu}(p,q) &=& {\cal S}_1\left (-g^{\mu\nu} + {q^\mu
                q^\nu\over q^2} \right ) + {\cal S}_2 {1\over m^2}
                \left (p^\mu - {m\nu\over q^2} q^\mu \right ) \left
                (p^\nu - {m\nu\over q^2} q^\nu \right ) + {\cal S}_4
                {1\over m^2} q^\mu q^\nu + {\cal S}_5 {1\over m^2}
                (p^\mu q^\nu + p^\nu q^\mu ) \nonumber \\ && -i{\cal
                A}_1{1\over m}\epsilon^{\mu\nu\alpha\beta}q_\alpha
                S_\beta - i\nu{\cal A}_2 {1\over m^2}
                \epsilon^{\mu\nu\alpha\beta} q_\alpha\left (S_\beta -
                {S\cdot q\over m\nu} p_\beta \right ) - i{\cal
                A}_3{1\over m}\epsilon^{\mu\nu\alpha\beta}p_\alpha
                S_\beta
\label{Tcore1}
\end{eqnarray}
with
\begin{eqnarray}
   {\cal S}_1(q^2,\nu) &=& -H_1(q^2,\nu) = -{q^2\over m^2}\overline
            H_2(q^2,\nu) - {\nu\over m}\overline H_4 (q^2,\nu) -
            {1\over 2m} a(q^2,\nu), \label{S1} 
\\ 
{\cal S}_2(q^2,\nu)
            &=& H_3(q^2,\nu) = {q^2\over m\nu} \overline H_4(q^2,\nu)
            + {1\over \nu} b(q^2,\nu), \label{S2} 
\\ 
{\cal S}_4(q^2,\nu) &=& {m\over 2 q^2} a(q^2,\nu) - {m^2\nu\over q^4}
            b(q^2,\nu), \label{S4} 
\\ 
{\cal S}_5(q^2,\nu) &=& {m\over q^2} b(q^2,\nu), \label{S5} 
\\ 
{\cal A}_1(q^2,\nu) &=& H_6(q^2,\nu) + {\nu\over m} H_8(q^2,\nu),
            \label{A1}
\\ 
{\cal A}_2(q^2,\nu) &=& \overline H_7(q^2,\nu) - H_8(q^2,\nu),
            \label{A2}
\\ 
{\cal A}_3(q^2,\nu) &=& {\nu\over m}c(q^2,\nu),\label{A3}
\end{eqnarray}
where $a(q^2,\nu)$, $b(q^2,\nu)$ and $c(q^2,\nu)$ depend on $q^2$ slowly.

The total virtual $\gamma^* N$ cross section is related to the forward
Compton scattering amplitude by the Optical theorem. For a specific
polarization of the incoming (virtual) photon or $Z$ particle
with helicity $\lambda$, it is written in the form of
\begin{equation}
   \sigma_{tot}^\lambda(\gamma^*N\to X) = {4\pi^2
             \alpha_{em}\over m K}
             \varepsilon^{\mu*}_\lambda \varepsilon^\nu_\lambda
             W_{\mu\nu},
\end{equation}
where $q_\mu\varepsilon^\mu_\lambda = 0$ ($\lambda=0,\pm$) and $K$ as
the flux factor, which is not unique when $Q^2\ne 0$, is chosen to be
$\sqrt{\nu^2-Q^2}$. $W^{\mu\nu}$ is defined as
\begin{equation}
    W^{\mu\nu} = {1\over 4\pi} \int d^4 x e^{iq\cdot x}\bra{pS}\left 
                 [ J^\mu_{core}(x),J^\nu_{core}(0) \right ]\ket{pS}.
\end{equation}
It has the same Lorentz structure as $T^{\mu\nu}_{core}$, namely,
\begin{eqnarray}
  W^{\mu\nu}(p,q) &=& W_1\left (-g^{\mu\nu} + {q^\mu q^\nu\over q^2}
                \right ) + W_2 {1\over m^2} \left (p^\mu - {m\nu\over
                q^2} q^\mu \right ) \left (p^\nu - {m\nu\over q^2}
                q^\nu \right ) + W_4 {1\over m^2} q^\mu q^\nu + W_5
                {1\over m^2} (p^\mu q^\nu + p^\nu q^\mu ) \nonumber 
\\
                && -iZ_1{1\over m}\epsilon^{\mu\nu\alpha\beta}q_\alpha
                S_\beta - i\nu Z_2 {1\over m^2}
                \epsilon^{\mu\nu\alpha\beta} q_\alpha\left (S_\beta -
                {S\cdot q\over m\nu} p_\beta \right ) - iZ_3{1\over
                m}\epsilon^{\mu\nu\alpha\beta}p_\alpha S_\beta
\label{Wcore1}
\end{eqnarray}
Optical theorem relates the imaginary part of the invariant Compton
amplitudes to the structure functions in the DIS
\begin{eqnarray}
W_{1,2,4,5}(q^2,\nu) &=& {1\over 2\pi} \mbox{Im} {\cal
S}_{1,2,4,5}(q^2,\nu+i0^+),\label{W-S}
\\ 
Z_{1,2,3}(q^2,\nu) &=& {1\over2\pi} \mbox{Im} {\cal
A}_{1,2,3}(q^2,\nu+i0^+).\label{Z-A}
\end{eqnarray}

\subsubsection{Kinematics and Observables}

The laboratory kinematics is defined in the usual way. The plane
formed by the 3-momenta of the incident and scattering leptons is
called the scattering plane.  The direction defined by
$\mbox{\boldmath{$q$}}$ is called the 3rd (or z) axis; the direction
perpendicular to the the scattering plane is called the 2nd (or y)
axis; the direction normal to the plane formed by the 2nd and 3rd axis
is called the 1st (or x) axis. The positive quantity $Q^2=-q^2$ is
used in the following discussion.

The virtual photon 4-momentum is
$q^\mu=\{\nu,0,0,\sqrt{\nu^2+Q^2}\}$. The polarization vector with
property $\varepsilon^{*\mu}_\lambda \varepsilon_{\lambda\mu}=1$ and
$q\cdot \varepsilon = 0$ are
\begin{eqnarray}
    \varepsilon_\pm^\mu = \mp {1\over\sqrt{2}}\{0,1,\pm i, 0\} &\hspace{0.2in}&
    \varepsilon_0^\mu = {1\over\sqrt{Q^2}}\{\sqrt{\nu^2+Q^2},0,0,\nu\}.
\end{eqnarray}
There is a fourth polarization state of the photon in the spontaneous
EM gauge symmetry breaking phase.  This state is some times called the
longitudinal mode in the literature. The longitudinal polarization is
proportional to $q^\mu$, namely,
\begin{equation}
    \varepsilon_s^\mu = {1\over\sqrt{-Q^2}} q^\mu.
\end{equation}
It does not contribute to the semi-leptonic 
processes interested in this paper.

The unpolarized total virtual $\gamma^* N$ cross sections are
\begin{eqnarray}
  \sigma_T &=& {1\over 2} \left ( \sigma_{tot}^+ + \sigma_{tot}^-
            \right ) = {4\pi^2\alpha_{em}\over m}
            {1\over K} W_1, \label{sigmaT} 
\\ 
  \sigma_L &=& \sigma^0_{tot} ={4\pi^2\alpha_{em}\over m}
            {1\over K} \left [\left ( 1 + {\nu^2\over
            Q^2} \right ) W_2 - W_1 \right ]. \label{sigmaL}
\end{eqnarray}
The difference between polarized ones with the photon and nucleon spin
polarization antiparallel and parallel respectively is
\begin{equation}
   \sigma_{TT} = {1\over 2}\left (\sigma_{1/2} - \sigma_{3/2} \right )
        = {4\pi^2\alpha_{em}\over m^2} {\nu\over K} \left [Z_1 - {Q^2\over
             m\nu}Z_2 + Z_c \right ],
\label{Pol-Sig1}
\end{equation}
and the interference cross section between ``longitudinal'' and
transverse photon polarization that is perpendicular to the scattering
plane is
\begin{equation}
   \sigma_{LT} = {4\pi^2\alpha_{em}\over m^2}
   {\sqrt{Q^2}\over K}\left [ Z_1 +
       {\nu\over m} Z_2 - {\nu^2\over Q^2} Z_c \right ],
\label{Pol-Sig2}
\end{equation}
where 
\begin{equation}
    Z_c = {m\over \nu}Z_3. \label{ZcDef}
\end{equation}

Here
\begin{eqnarray}
   W_a(-Q^2,\nu) &=& {1\over 2\pi} \mbox{Im} a(-Q^2,\nu+i0^+),\label{Wadef}\\
   W_b(-Q^2,\nu) &=& {1\over 2\pi} \mbox{Im} b(-Q^2,\nu+i0^+),\label{Wbdef}\\
   Z_c(-Q^2,\nu) &=& {1\over 2\pi} \mbox{Im} c(-Q^2,\nu+i0^+).\label{Zcdef}
\end{eqnarray}

The laboratory frame inclusive $eN$ or $\mu N$ unpolarized scattering cross
section are related to these $\gamma^*$ total cross section
\begin{eqnarray}
   {d\sigma\over dE'd\Omega} &=& \Gamma (\sigma_T + \epsilon \sigma_L
      ) \nonumber 
\\ 
    &=& {\alpha_{em}^2\over 8mE^2 \sin^4\theta/2}
      \left [ \left ( 2 + 2\sin^2\theta/2 + {y^2\over 1-y}\right ) W_2
      - {y^2\over 1-y} W_L \right ],
\label{dsigmalN}
\end{eqnarray}
where $y = P\cdot q/P\cdot k$ with k the 4-momentum of the initial
lepton, for the unpolarized cross section. Here
\begin{eqnarray}
    \epsilon &=& \left [ 1+2\left (1+{\nu^2\over Q^2} \right ) \tan^2
                 {\theta\over 2} \right ]^{-1},\label{eps-param} 
\\
                 \Gamma &=& {\alpha_{em}\over 2\pi^2} {E'\over E}
                 { K\over Q^2} {1\over
                 1-\epsilon}.\label{Gamma-param}
\end{eqnarray}

   If the nucleon is polarized along $\mbox{\boldmath{$q$}}$
\begin{equation}
  \left ( {d\sigma_{-1/2}\over dE'd\Omega} - {d\sigma_{+1/2}\over
  dE'd\Omega} \right ) = 2\Gamma \sqrt{1-\epsilon^2}
         \sigma_{TT}.
\label{dsigmalN-p1}
\end{equation}
If the nucleon is polarized perpendicular to $\mbox{\boldmath{$q$}}$ and
parallel to the scattering plane
\begin{equation}
  \left ( {d\sigma_{-1/2}\over dE'd\Omega} - {d\sigma_{+1/2}\over
  dE'd\Omega}\right ) = 2\Gamma
  \sqrt{2\epsilon(1-\epsilon)}\sigma_{LT}.
\label{dsigmalN-p2}
\end{equation}

\subsubsection{The two pomeron hypothesis}

The presence of finite $a$, $b$ and $c$ is not a sufficient condition
for the spontaneous partial breaking of the EM gauge symmetry. In
fact, the same form of observable quantities in Eqs.\ref{Tcore1} and
\ref{Wcore1} with additional terms $\{{\cal S}_4,{\cal S}_5,{\cal A}_3
\}$ and $\{W_4,W_5,Z_3\}$ can be derived from a theory in which a
global $U(1)$ symmetry that is contained in the EM local gauge symmetry
is explicitly broken instead of spontaneously. The difference between
them lies in more details related to the existence of extra strong
interaction channel in the semi-leptonic processes discussed in
subsection \ref{sec:general-pic}.

\begin{figure}[h]
\begin{center}
\epsfbox{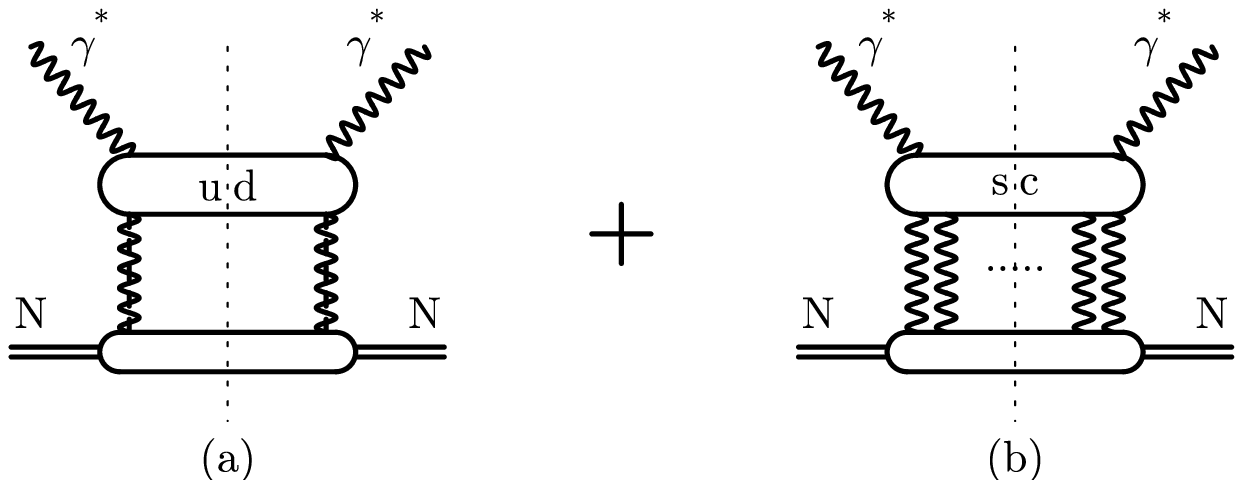}
\end{center}
\caption{\label{Fig:MultiPart} Forward Compton scattering amplitude
diagrams that contribute to the multi-particle production. For the EM
contribution to the $NN$ scattering, Figures (a) and (b) are
comparable in their energy dependence. In the semi-leptonic $lN$
interaction, figure (a) contains the normal contributions and figure
(b) contains extra interaction components due to the (lack of the)
Goldstone boson. We shall assume the Goldstone boson can be reggonized
to become the ``hard pomeron''. }
\end{figure}

The inclusive cross section in a semi-leptonic process are related to
the imaginary part of the forward virtual photon and nucleon Compton
scattering amplitude which is schematically shown in
Fig. \ref{Fig:MultiPart}. Let us view this amplitude as the
self-energy of the photon when propagate inside a nucleon. If the
virtual photon is emitted by a nucleon, then there is a companion
Goldstone boson with the proper strength that travel with the
photon. Due to the Higgs mechanism, these two companion excitations
effectively screen each other by canceling each others strong
interaction components discussed in subsection \ref{sec:general-pic}
so that besides other strongly interacting particles like gluons and
quark--antiquark pairs, the subsequent photons and Goldstone bosons
that are generated by them are also of proper portion that they cancel
each other's strong interaction effects. On the other hand, if the
virtual photon is emitted by a lepton, then there is no companion
Goldstone boson, the passage of this virtual photon will meet greater
resistance because its effective interaction with the environment
inside the nucleon is in the strong interaction range. It also would
frustrate the portion of the vacuum state inside the nucleon so that a
disproportional number of Goldstone are generated, which in turn
generate more final hadronic states after they are hadronized.

The hadronization of the Goldstone bosons (or the lack of it, see
Eq. \ref{GTstrng}) are also selective. They would not hadronize into
hadrons made up of up and down quarks, like the nucleons, $\rho$
mesons, etc., since the coupling between the Goldstone boson is always
screened by the virtual photon inside the system. Their coupling to
other hadrons made up of quarks that do not participate in the
symmetry breaking. These quarks are strange quarks, charm and bottom
quarks, etc. according to the simplest scenario of color
superconductivity at low density in
Refs. \cite{cspc1,cspc2,cspc3,scalar-spc,Wilczek,Shuryak}. Therefore
the difference between the total cross section of the $NN$ scattering
and the semi-leptonic $lN$ scattering should lie in the difference in
the contributions of the strange and charm/bottom final state like the
$\phi$ meson, $J/\Psi$ mesons, etc., not in the regular hadronic
states made of up and down quarks, like the $\rho$, $\omega$, $\pi$
mesons.

 The partial spontaneous breaking of the EM gauge symmetry inside of a
nucleon is characterized by three non-vanishing pole strength of the
longitudinal components of the scattering amplitude
Eqs. \ref{H2decomp}--\ref{H7decomp} for {\em semi-leptonic processes}.
Since these pole terms are absent for the EM interaction part of
nucleon--nucleon scattering due to the realization of the Higgs
mechanism in such a system, it would be natural to assume that the effects
of the extra particles produced in a semi-leptonic $lN$ scattering
compared to the $NN$ scattering are encoded in $a(q^2,\nu)$,
$b(q^2,\nu)$ and $c(q^2,\nu)$, which should increase with the photon
energy faster than that are allowed by the Froissart bound valid for
the $NN$ interaction.

According to the phenomenological success of the Regge asymptotics, the high 
energy behavior of $H_1$,$\overline H_2$, $H_3$, $\overline H_4$, $H_5$,
$H_6$, $\overline H_7$ and $H_8$ follows that of the Regge asymptotics with
the soft pomeron as the leading trajectory. It is known that the soft pomeron
has an intercept of $\alpha_{\cal P}\sim 1$.

 For the believers of two pomerons, phenomenology \cite{Pomeron1}
implies that $\sigma_{\gamma^* N} \sim \nu^{\alpha_{{\cal P}'-1}}$,
with $\alpha_{{\cal P}'}$ the intercept of the ``hard pomeron'' of
order 1.4, for sufficiently large $Q^2$. Let us take the
two pomeron hypothesis in the present context and derive its consequences
in the following. 

In the Regge theory, it would be natural to assume
that the high energy behavior of the leading singularity term $W_b$ to
have the following high energy behavior
\begin{equation}
   W_b \sim \nu^{\alpha_{{\cal P}'}-1}.
\label{Wb-hi-E}
\end{equation}
The high energy behavior of $W_a$ is not known from experimental
observations due to the domination of $W_b$ in the cross section. By
assuming the universality of the Regge asymptotics even in the case of
hard pomeron, one can write
\begin{equation}
   W_a \sim \nu^{\alpha_{{\cal P}'}-1}.
\label{Wa-hi-E}
\end{equation}
Although there is no strong reason in doing so. $W_b$ is an isoscalar 
and $W_a$ contains isoscalar and isovector component. The high energy
behavior of $W_a$ can in principle be extracted from the data.
But as a working hypothesis, I shall assume that the isovector component of
$W_a$ also satisfy Eq. \ref{Wa-hi-E}.

 The $Z_1+Z_c$ term is related to the amplitudes for $\gamma^*\gamma^*\to N
\overline N$ with the two transversely polarized photon having opposite 
helicity in the center of mass frame. The isoscalar component of each
of them have the following asymptotic behavior at hight energies in
the $\gamma^* N \to \gamma^* N$ channel, namely,
\begin{eqnarray}
    Z_{1S} &\sim &\nu^{\alpha_{{\cal P}}-2},\label{Z1S} \\
    Z_{cS} &\sim &\nu^{\alpha_{{\cal P}'}-2},\label{ZcS}
\end{eqnarray}
where $\alpha_{\cal P}$ is the intercept of the soft
pomeron and the leading trajectory for $Z_c$ is assumed to be that of
the hard pomeron also.  There is no general argumentation which allows
us to write down the high energy behavior for their isovector
components.

   In principle, the energy dependence of $a$, $b$ and $c$ and their
imaginary parts may have a completely different $\nu$ dependence. For
fixed $Q^2$, assuming the asymptotic behavior Eqs. \ref{Wb-hi-E},
\ref{Wa-hi-E} and \ref{ZcS} hold, crossing symmetry requires that the
corresponding amplitudes are of the following general form
\begin{eqnarray}
  a(\nu) &=& P_a(\nu) + {2\over\pi}\int_{\nu_{th}}^\infty 
           {\nu'\mbox{Im} a(\nu')\over
                        {\nu'}^2-\nu^2} d\nu',\label{a-disp}\\
  b(\nu) &=& P_b(\nu) + {2\nu\over\pi}\int_{\nu_{th}}^\infty 
                      {\mbox{Im} b(\nu')\over
                        {\nu'}^2-\nu^2} d\nu',\label{b-disp}\\
  c(\nu) &=& P_c(\nu) + {2\over\pi}\int_{\nu_{th}}^\infty 
                       {\nu'\mbox{Im} c(\nu')\over
                        {\nu'}^2-\nu^2} d\nu',\label{c-disp}
\end{eqnarray}
where $P_{a,b,c}(\nu)$ are polynomials of $\nu$. The $Q^2$ dependence
in the above equations are suppressed. $P_c(\nu)=\mbox{const}$ and
possibly non-vanishing is known (see the following discussion) from
the study of the modification of GDH sum rule \cite{GDHpap} motivated
by analysis of experimental data. The form of $P_{a,b}$ remains to be
determined.

The isospin structure of $a$, $b$ and $c$ and their imaginary parts
can be found under the scenario
of Refs. \cite{cspc1,scalar-spc,Wilczek} and \cite{GDHpap}.  Parameter
$b$ is the strength of the double pole in $q^2$ in the Compton
scattering amplitudes; it corresponds to the last term of
Eq. \ref{Nbbdecomp}. Since the broken charge is an isoscalar, $b$ is
also an isoscalar. Parameter $a$ and $c$ are the strength of the
single pole in $q^2$ in the Compton scattering amplitude, they contain
contributions from the second and third terms in Eq. \ref{Nbbdecomp}
and the second term in Eqs. \ref{IsoOdd1} and \ref{IsoOdd2}. Therefore
\begin{eqnarray}
   a = a_S+a_{V} &\hspace{1in}&  W_a = W_{aS} + W_{aV} \label{EQ1}\\
   b = b_S\phantom{+a_{VS}}  &\hspace{1in}&  W_b = W_{bS} \label{EQ2}\\
   c = c_S+c_{V} &\hspace{1in}&  W_c = W_{cS} + W_{cV} \label{EQ3}
\end{eqnarray}
So, the most singular piece of the Compton amplitude in $q^2$ can be
observed in its isoscalar component and the next singular pieces of
the Compton amplitude manifested themselves in the isovector component
of its amplitude.

\section{The $Q^2\to 0$ limit and small $Q^2$ region}
\label{sec:Q20}

Let us assume that all of the structure functions considered are
regular in $Q^2$ when $Q^2$ is small, then following leading $1/Q^2$
behavior at small $Q^2$ hold
\begin{eqnarray}
    W_2 &\to & {1\over\nu} W_b,\label{W2asmp}\\
    W_L &\equiv & -{Q^2\over\nu^2} W_1 + \left (1+{Q^2\over\nu^2} \right ) W_2 
          \to {1\over\nu} \left [W_b + {Q^2\over \nu^2}
           \left (W_b + {\nu\over 2m} W_a \right )\right ], \label{WLasmp}
\end{eqnarray}
which incorporate the possibility of spontaneous partial breaking of
the EM gauge symmetry by assuming a non-vanishing $W_{a,b}$.
Here all terms of order $Q^2$ or higher are dropped except those ones
that depend on $W_a$ and $W_b$ since they have a quite different
energy dependence from the normal ones.

If $W_{a,b}\ne 0$ and $Z_c\ne 0$, then the following limiting
behavior follows
\begin{eqnarray}
    \sigma_L &\to& {4\pi^2\alpha_{em}\over m}{1\over K}{1\over 2 m}
           \left [W_a + {2m\nu\over Q^2} W_b + {2m\over\nu} W_b
     \right ],\label{sigmaLasymp} \\
    \sigma_{LT} &\to& 
         {4\pi^2\alpha_{em}\over m^2} {\sqrt{Q^2}\over K}
           \left [ -{\nu^2\over Q^2} Z_c \right ], \label{sigmaLTasymp}
\end{eqnarray} 
when $Q^2$ is small.

\subsection{Finite size effects}

The longitudinal total virtual photon--nucleon scattering cross
sections $\sigma_L$ and $\sigma_{LT}$ behave like $1/Q^2$ according to
Eqs. \ref{sigmaLasymp} and \ref{sigmaLTasymp}. Such a behavior can not
continue all the way down to $Q^2=0$ since a nucleon is a finite
system, such a singular behavior can not be fully developed in a
finite system. The rise of the cross sections as $Q^2\to 0$ is
saturated at certain small $Q^2 = Q^2_0$ value with $Q_0^2$ decreasing
function of the photon energy $\nu$ \cite{GDHpap}.  The simplest way
to represents the finite size (of the nucleon) effects is to replace
$1/Q^2$ in the pole terms by the following ad hoc form
\begin{equation}
      {1\over Q^2} \to {1\over Q^2+Q_0^2},\label{Q2smear}
\end{equation}
for the definiteness of the discussion. It turns to $1/Q_0^2$ instead
of infinity in the $Q^2\to 0$ limit. In addition, according to the
discussions of Ref. \cite{GDHpap}, the maximum of the peak at the
$Q^2=0$ position is proportional to the maximum number of the coherent
interaction of the photon with the finite medium at large $\nu$. So we
can further choose a specific form for $Q_0^2$ 
\begin{equation}
     Q^2_0 = {1\over A + B\nu} \label{Q0}
\end{equation}
as a working hypothesis with parameters $A$ and $B$ slow varying function of $\nu$. It must
be emphasized that the functional forms given in Eqs. \ref{Q2smear} and
\ref{Q0} are chosen only based on qualitative
considerations. Other forms having the same qualitative properties may
be better when detailed fitting to the experimental data are made.

  Therefore, in the $Q^2\to 0$ limit, one should replace $a$, $b$ and $c$ by
\begin{eqnarray}
    a &\to & {Q^2\over Q^2 + Q_0^2} a,\label{a-repl}\\
    b &\to & {Q^4\over (Q^2 + Q_0^2)^2} b,\label{b-repl}\\
    c &\to & {Q^2\over Q^2 + Q_0^2} c,\label{c-repl}
\end{eqnarray}
and their imaginary part by
\begin{eqnarray}
    W_a &\to & {Q^2\over Q^2 + Q_0^2}W_a,\label{Wa-repl}\\
    W_b &\to & {Q^4\over (Q^2 + Q_0^2)^2} W_b,\label{Wb-repl}\\
    Z_c &\to & {Q^2\over Q^2 + Q_0^2} Z_c.\label{Wc-repl}
\end{eqnarray}

\subsection{The real photon limit}

\subsubsection{High energy behavior}

After the substitutions Eqs. \ref{a-repl}--\ref{Wc-repl} are made in
Eqs. \ref{dsigmalN}, \ref{dsigmalN-p1} and \ref{dsigmalN-p2}, it is
easily seen that terms involving $W_{a,b}$ and $Z_c$ are actually
higher order terms in $Q^2$ than the leading term. The unpolarized
$\gamma^* N$ cross section is
\begin{equation}
    \sigma_{\gamma N} = \lim_{Q^2\to 0} (\sigma_T + \epsilon \sigma_L )
         = {4\pi^2\alpha_{em}\over m^2} \sigma_0 (0,\nu)
\end{equation}
with
\begin{equation}
    \sigma_0(Q^2,\nu) = - {1\over 2\pi} \mbox{Im} \overline H_4(Q^2,\nu).
\end{equation}
Therefore there seems to be no anomalies here. The real
photon--nucleon total cross section has the same form as the one in
which the EM gauge symmetry is not broken. The high energy behavior of
$\overline H_4$ obtained from $H_4$ by subtracting the possible pole
term in $Q^2$ is expected to follow the Regge asymptotics: $\overline
H_4 \sim \nu^{\alpha_{\cal P}-1}$ with the leading trajectory that of
the soft pomeron having an intercept $\alpha_{\cal P}=1.0808$. By
assuming that the imaginary part of $\overline H_4$ has the same large
energy asymptotics, one get
\begin{equation}
   \sigma_{\gamma N} \sim \nu^{\alpha_{\cal P}-1} = \nu^{0.0808}.
\end{equation}
It agrees with the experimental observation \cite{Abram} well.

  The polarized $\gamma N$ cross section are
\begin{eqnarray}
  {1\over 2} (\sigma_{1/2} - \sigma_{3/2} ) &=&
  {4\pi^2\alpha_{em}\over m^2} (Z_1 +Z_c),
\label{SigP0} \\
   \sigma_{LT} &=& 0.
\label{SigLT0}
\end{eqnarray}

\subsubsection{The modification of the GDH sum rule}

The first of the above equations is studied in the context of the
apparent violation of the GDH sum rule \cite{GDHpap}, which states
\begin{equation}
    \int_{\nu_{th}}^\infty {\sigma_{1/2}(\nu)-\sigma_{3/2}(\nu)\over \nu} d\nu 
                  = {2\pi^2\alpha_{em}\over m^2} \kappa^2
\end{equation}
with $\kappa$ the nucleon anomalous magnetic moment and $\nu_{th}$ the
inelastic threshold in $\nu$.  If the EM gauge symmetry is
spontaneously partial broken, then a modification of this sum rule
\cite{GDHpap} is
\begin{equation}
    \int_{\nu_{th}}^\infty {\sigma_{1/2}(\nu)-\sigma_{3/2}(\nu)\over \nu} d\nu 
                  = {2\pi^2\alpha_{em}\over m^2} \left (\kappa^2+2c_\infty \right )
\end{equation}
in the notation of this paper. The parameter $\rho_\infty$ defined in
\cite{GDHpap} is related to the polynomial $P_c(\nu)$ in
Eq. \ref{c-disp} in the following way
\begin{equation}
   \rho_\infty = {1\over m^2} c_\infty = {1\over m^2} \lim_{Q^2\to
   0}\lim_{\nu\to\infty} c(Q^2,\nu)
\end{equation}
where $c_\infty=\lim_{\nu\to\infty} P_c(\nu)$. 
So finite modification of the GDH sum rule constrained by observation
restricts the polynomial $P_c(\nu)$ to be at most a finite constant
$c_\infty$. 

\subsubsection{Comments}

\begin{enumerate}
\item
The current analysis of published data \cite{GDHt}
indicates that $c_\infty\ne 0$ and mostly isovector. The most recent
experimental study of GDH sum rule at MAMI in Mainz, Germany reduced
the discrepancy between the theory and data to within $10\%$ for a
proton based on a preliminary analysis of the data taken. Similar
study on a neutron, which is relevant to the isovector component of
the difference that can even be a violation in sign, is still lacking.
The theoretical extrapolation of the DIS data \cite{GDHt2}, which
includes all possible photon energies also has a less than $10\%$
difference for the GDH sum rule. Therefore whether or not the
GDH sum rule is in fact violated is still an open question.

\item

Since according to Eq. \ref{EQ3}, c contains both the isoscalar and
isovector components, the fact that the isoscalar component gets
suppressed can also be qualitatively understood. This is because the
isoscalar component relates to the second and third terms in
Eq. \ref{Nbbdecomp}, which contain $b_{core}^\mu$ that are reduced in
strength compared to its value in the normal phase in which EM gauge
symmetry is kept. On the other hand, the isovector component of $c$
relates to the second terms in Eqs. \ref{IsoOdd1} and \ref{IsoOdd2},
which do not contain $b_{core}^\mu$ and therefore is not suppressed.

\item

 Before continuing, a discussion regarding the subtle nature of $Z_c$
term in Eq. \ref{SigP0} is necessary.  Due to the separation of the
pole terms from the invariant functions $H_i$ ($i=1,2,\ldots,8$) in
the forward Compton scattering amplitudes, the intermediate results in
deriving the modified GDH sum rule is slightly different from those in
Ref. \cite{GDHpap}. Here ${\cal A}_1+c$ corresponds to $mA_1$ of
Ref. \cite{GDHpap}.  The dominant contribution to the left hand side
of the GDH sum rule is from the photo pion production through the
$\Delta$ resonance. $Z_c$ does not contain such contributions
however. The reason is the pion and any other mesons made of up and
down quarks belong to the hadronic subspace in which the Higgs
mechanism fully operates. Since $Z_c$ is the contribution of the
Goldstone boson, it is only significant beyond the $\phi$ production
threshold.  The $\phi$ and $J/\Psi$ contains no up or down quarks so
that they do not couple to the Goldstone boson.  Since the Goldstone
boson couples to the hadronic matter with a strength of strong
interaction, the lack of the contribution of the Goldstone boson means
that the EM production of $\phi$ and $J/\Psi$ contain an uncanceled
extra component with the rate comparable to that of the strong
interaction if the EM gauge symmetry is spontaneously partial
broken. Therefore the effect of $Z_c$ on the left hand side of GDH sum
rule can be ignored below the $\phi$ production threshold, which is
greater than 1 GeV, because it is of higher order in
$\alpha_{em}$. The contribution of these states to the sum rule are
expected to be small due to the high threshold and the suppression
imposed by Okubo--Zweig--Iuzuka (OZI) rule \cite{OZI}.

\end{enumerate}

\subsection{The small $Q^2$ region}

 The small $Q^2$ region is defined as the region in which the pole
behaviors of $a$, $b$ and $c$ is not dominant, namely, the region in
which $Q^2$ is not much larger than $Q_0^2$. This is a
transition region in which the effects of the symmetry breaking terms $a$,
$b$ and $c$ and all other normal terms compete with each other due to
the finite size effects of a nucleon.

The more detailed description of this region is beyond the scope of
this work, which is based mainly on the symmetry considerations. This
is a region that can be experimentally studied in a more quantitative
way.

\section{The deep inelastic, small $x$ region}
\label{sec:DIS}

\subsection{Inclusive processes}

 In the Bjorken limit defined as the kinematic region in which $Q^2 =
-q^2 \to \infty$, $\nu\to\infty$ and $x=Q^2/2m\nu=$fixed, the quark
substructure of a nucleon begin to reveal itself in the structure
functions of the nucleon by exhibiting scaling. This is because the
large $Q^2$ virtual photon $\gamma^*$ interacts with a quark inside a
nucleon with a (light cone) time duration much smaller than the soft
and coherent processes that symmetry breaking and confinement of
quarks take place. Thus the virtual photon provides a snapshot of the
quark distribution of a nucleon on the light cone. Combined with
the asymptotic freedom, this leads to 
the parton model in which a factorization of the leading twist hard
processes and the much slower soft processes can be factorized.  In the
region where $Q^2$ is much smaller than $\nu$, which means small $x$,
the light cone time duration in the current--current correlator
becomes much larger than the virtual photon's wave length, the
coherent processes can accumulate their strength
\cite{GDHpap}. Therefore one expects 1) the manifestation of the
leading twist coherent subprocesses due to symmetry breaking, if any,
in these high energy reactions and 2) non-leading twist contributions.

The scaling properties of the structure functions in the Bjorken
limit make it useful to write them as
\begin{eqnarray}
 F_1(x,Q^2) &=& W_1(-Q^2,\nu), \label{F1}\\
 F_{2,4,5}(x,Q^2) &=& {\nu\over m} W_{2,4,5}(-Q^2,\nu), \label{F245}\\
 F_L(x,Q^2) &=&  {\nu\over m}W_L(-Q^2,\nu) = F_2(x,Q^2)-2xF_1(x,Q^2),\label{FL}\\
 g_1(x,Q^2) &=& {\nu\over m} \left [ Z_1(-Q^2,\nu) + Z_c(-Q^2,\nu) \right ],\label{G1}\\
 g_2(x,Q^2) &=& {\nu^2\over m^2} Z_2(-Q^2,\nu).\label{G2}
\end{eqnarray}
True scaling means that these structure functions are independent of
$Q^2$ at large enough $x$.  In QCD, they change only slowly as a
function of $Q^2$. Such a scaling behavior of the above defined
structure functions has a rather natural interpretation in terms of
parton model. 

In addition to these conventional structure functions,
a set of new ones characterizing the symmetry breaking are defined in
the following
\begin{eqnarray}
 F_{a,b}(x,Q^2) &=& W_{a,b}(-Q^2,\nu),\label{Fab}\\
 g_c(x,Q^2) &=& {\nu\over m} Z_3(-Q^2,\nu).\label{G3}
\end{eqnarray}
Since, as it is shown above, the physical processes interested here
due to the spontaneous partial breaking of the EM gauge symmetry {\em
does not depend on a particular finite mass scale}, it is expected
that scaling behavior should be present in certain forms at high
energies.

{\em It is assumed that these new structure functions change slowly
with $Q^2$ in a logarithmic fashion in the paper for the further
development of the ideas that can be experimentally
examined}\footnote{Note that the possible existence of a virtual phase for
the strong interaction vacuum that spontaneous partial breaks the EM
gauge symmetry is a {\em necessary} condition for making such an
assumption, but it is not a sufficient one.}.

Under such an assumption,
it is expected that they are only significant in the small $x$
kinematic region as functions of $x$ in which coherent processes
starts to build up.

 The unpolarized differential cross section for the inclusive
processes is
\begin{equation}
{d^2\sigma\over dx dQ^2} = {2\pi\alpha_{em}^2\over x Q^4} \left \{ \left [
      1+ (1-y)^2 \right ] F_2(x,Q^2) - y^2 F_L(x,Q^2) \right \}.
\label{DISunpol}
\end{equation}

The polarized differential cross section for the inclusive process proportional
to the polarized total virtual photon and nucleon cross sections in Eqs.
\ref{dsigmalN-p1} and \ref{dsigmalN-p2}. $\sigma_{TT}$ and $\sigma_{LT}$ 
are related to the spin structure functions $g_1$ and $g_2$ in the following
way
\begin{eqnarray}
  \sigma_{TT} &=& {4\pi^2\alpha_{em}\over m K} \left (g_1 -
              {Q^2\over \nu^2} g_2 \right ), \label{sTTg1g2}\\
  \sigma_{LT} &=& {4\pi^2\alpha_{em}\over m K} {\sqrt{Q^2}\over\nu}
                  \left [g_1 + g_2 - 
      \left ( {m\over\nu} + {1\over 2 x} \right ) g_c  \right ]
\label{sLTg1g2}
\end{eqnarray}
with 
\begin{equation}
    g_c = {\nu^2\over m^2} Z_c. 
\end{equation}

Due to the presence of the quite singular term $g_c$ in
Eq. \ref{sLTg1g2}, if symmetry breaking allows for $g_c\ne 0$, then the
structure functions $g_1$ and $g_2$ at small $x$ can not be
extracted from the observables
\begin{eqnarray}
    A_1 &=& {\sigma_{TT}\over \sigma_T}, \\
    A_2 &=& {\sigma_{LT}\over \sigma_T}.
\end{eqnarray}
Instead, only the following combination can be extracted
\begin{eqnarray}
   \tilde g_1 &=& g_1 - { Q^2\over \nu^2+Q^2} \left ( {m\over \nu} 
        + {1\over 2x} \right ) g_c 
              = {\nu^2\over \nu^2+Q^2} F_1
              \left (A_1 + {\sqrt{Q^2}\over \nu} A_2 \right ),
\label{tildeg1} \\
   \tilde g_2 &=& g_2 - {\nu^2\over \nu^2+Q^2} \left (
         {m\over \nu} + {1\over 2x} \right ) g_c
      = {\nu^2 \over \nu^2+Q^2} F_1 \left (-A_1+{\nu\over \sqrt{Q^2}} A_2 
  \right )
\label{tildeg2}
\end{eqnarray}

In the Bjorken limit, we have
\begin{eqnarray}
   \tilde g_1 &=& g_1  
              =  F_1 A_1, 
\label{g1-Blmt}\\
   \tilde g_2 &=& g_2 - {1\over 2x} g_c
      =  {\nu\over \sqrt{Q^2}} F_1 A_2. 
\label{g2-Blmt}
\end{eqnarray}
Namely, it is impossible to extract $g_2$ at small $x$ 
from the data on $A_1$ and $A_2$ even
in the true Bjorken limit. 

\subsection{Diffractive meson production}

The diffractive meson production in the DIS processes provide
additional means for the check various theoretical pictures for the
DIS at small $x$. The pomeron exchange picture is proven to be a good one to
describe the data \cite{Diffr}.  The testable signature for the
scenario proposed in this work are the following
\begin{enumerate}
\item
According to the scenario proposed in this work, the diffractive
production of the $\rho$ meson, which is made up of up and down
quarks, is dominated by the soft pomeron with an intercept of
$\alpha_{\cal P} \sim 1.08$. This is because, as discussed above, the
Hilbert space spanned by the  Goldstone boson does not
contribute to the $\rho$ meson production.
\item
The semi-leptonic electro- diffractive production of the $\phi$ and
$J/\Psi$ mesons, on the other hand, is dominated by the hard pomeron
with an intercept of $\alpha_{{\cal P }'} \sim 1.4$ since the strange
and charm quarks, like the charged leptons, have zero nucleon
charge. So the cross section for the process is described by the
imaginary part of Fig. \ref{Fig:MultiPart}.b, which, due to the (lack
of) the contribution of the Goldstone boson in the up and down quark
subspace, couples to the nucleon with the strength of strong
interaction. The change of the diffractive cross section for the
$\phi$ and $J/\Psi$ production should follow that of the hard pomeron
if there is a spontaneous partial breaking of the EM gauge symmetry
according to the scenario proposed here.
\item
As indicated by Eq. \ref{jsprd-jcore}, the core part of the EM current
operator for the up and down quarks is not conserved in the symmetry
breaking case, it follows from Eq. \ref{Semi-lp-T} that 
current conservation shall {\em appear} to be violated in exclusive 
processes like the diffractive meson production. 
\end{enumerate}

\subsection{Comments on phenomenological observations}

\begin{enumerate}
\item

   If the rapid increase of measured $F_2$ is attributed to the
symmetry breaking terms $W_a$ and $W_b$, which are allowed to increase
beyond the Froissart bound like Eq. \ref{Wb-hi-E} and \ref{Wa-hi-E},
then we have observed behavior 
\begin{equation}
    F_2(x) \sim W_b \sim x^{1-\alpha_{{\cal P}'}} \approx x^{-0.4}
\end{equation}
at small $x$. It agrees with the hard Pomeron interpretation given here.

\item 
   The structure function $F_2(x)$ is extracted from the measured
unpolarized cross section corresponding to the left hand side of
Eq. \ref{DISunpol}. In order to get $F_2$, longitudinal structure
function $F_L$ at small $x$, which is believed to be small, has to be
known. The next-to-leading order perturbative QCD calculation of $F_L$
is shown to incorporate the data, but it could be a self-consistent
game \cite{Thorne} in the small $x$ region.  There could be other
values for $F_2$ and $F_L$ that also describe the data on the left
hand side of Eq. \ref{DISunpol}. Independent experimental
determination of $F_L$ is desirable.  For fixed initial lepton energy
experiment, in which the quantity $y$ is related to $Q^2$ by
$y=Q^2/sx$ with $s=2m\epsilon_l$ and $\epsilon_l$ the energy of the
incident charged lepton, it is difficult to measure $F_L$ since the
$Q^2$ dependences of both $F_2$ and $F_L$ are also a measured
quantities.
\item 
If the experiment can be designed to take data with fixed $Q^2$ and $x$, then 
the $y$ dependence of the cross section on the left hand side of
Eq. \ref{DISunpol}, which is only quadratic in $y$, can be used to
determine both $F_2$ and $F_L$ independently. Of special interest to
the scenario proposed in this work is the small $x$ region where $F_2$ 
is shown to rise quickly. If it is indeed that the symmetry
breaking terms $W_a$ and $W_b$ are causing the quick changes of $F_2$
at small $x$, then $F_2$ and $F_L$ at small enough $x$ will be
\begin{eqnarray}
    F_2 &\approx & {1\over m} W_b, \label{F2behv} \\
    F_L &\approx & {1\over m} \left ( W_b + x W_a \right ) \label{FLbehv}.
\end{eqnarray}
Let us further suppose that $W_b>> xW_a$ at small enough $x$, then
the right hand side of Eq. \ref{DISunpol} will be a linear rather
than quadratic function of $y$.

\item
  The violation of the Gottfried sum rule is related to the unexpected
behavior of the isospin odd component of $F_2$. It is believed to be
caused by the small $x$ region of $F_2(x)$. According to Eq.
\ref{F2behv}, the symmetry breaking effects, if present, do not
contribute to it since $W_b$ is an isospin even term (see
Eq. \ref{EQ2}).  But this idealized situation was not realized in the
analysis of the data because $F_L$ is in fact unknown from direct
observation. The extracted $\tilde F_2(x)$ under certain theoretical
assumption may contain small contribution of the true $F_L$, namely
\begin{equation}
   \tilde F_2 = F_2 + \eta F_L
\end{equation}
with $\eta$ a small number that can has a $x$ dependence.  Since $F_L$
contains $W_a$, which contains an isospin odd component (see
Eq. \ref{EQ1}), the extracted $\tilde F_2$ contains a term that
violate the Gottfried sum rule. But this violation is expected to
diminish when more information about $F_L$ at small $x$ is known
according to the scenario proposed here. Since from Eq. \ref{Wa-hi-E},
$F_L \sim x^{2-\alpha_{{\cal P}'}} \sim x^{0.6}$, the isospin odd
component of $\tilde F_2$ contains a small piece that decreases to
zero very slowly compared to the small $x$ extrapolation used to find
the discrepancy. This could be the source of the violation of the
Gottfried sum rule.
\begin{table}[h]
\caption{The effective charges for the electro-weak couplings in the
  DIS between leptons and a nucleon/nucleus 
   according to the standard
  model. $\tilde C_V$ and $\tilde C_A$ are coefficients of the vector
  and axial vector current operators 
  in the hadronic weak neutral current operator.
  Here $\theta_W$ is the Weinberg angle.  ``$u$'' and ``$d$''
  denote up and down quarks respectively. The corresponding 
  charge for an anti-quark is just opposite. The value for $\alpha$ in
  the table depends on the color and the momentum fraction $x$ 
  of the corresponding quark.
\label{Tab:compare}}
\begin{tabular}{|c|cccc|}
Quark&$\tilde Q$($l + h\to l+ h$) &$\tilde C_A$($\nu + h\to \nu + h$)
&$\tilde C_V$($\nu + h\to \nu + h$)&\phantom{$\displaystyle ab \over c$}
 \\
\hline
 &&&& \\
 \large{u} & $\displaystyle \alpha {1\over 6} + {1\over 2}$ & $\displaystyle
{1\over 2}$ & $\displaystyle{1\over 2} -
  ({\alpha\over 3} + 1) sin^2\theta_W $&\\
 &&&& \\
 \large{d} &  $\displaystyle\alpha {1\over 6} - {1\over 2}$ & $\displaystyle
-{1\over 2}$ & $\displaystyle-{1\over 2} -
   ({\alpha\over 3} - 1) sin^2\theta_W$ &\\
 &&&& 
\end{tabular}
\end{table}

\item

The difference between the charge lepton DIS data and the neutrino DIS
data at small $x$ ($x<0.1$) remains after heavy target corrections are
included \cite{MRST}. Such a difference is anticipated
\cite{PCAC}. The reason is because the isoscalar component of the EM
charge (the nucleon number density current) of the up and down quarks
contributes differently to the charge lepton neutral current DIS,
neutrino neutral current and neutrino charged current DIS. For the
charged lepton neutral current DIS, the isoscalar charge is $e/6$. It
is $-e \sin\theta_W/3$ for neutrino neutral current DIS, which is
smaller in magnitude (see Table \ref{Tab:compare}). Here, $\theta_W$
is the Weinberg angle. The isoscalar charge does not contribute to the
neutrino charged current DIS processes.

The structure function $F_2$ is extracted from the experimental data
based on the assumption that the core charge of quark is not
modified. So The reduction of the core charge of the light quarks in
the superconducting phase is expected to reduce the extracted
structure function against its true value.  Such a reduction is larger
for the results from charge lepton neutral current DIS than the one
from neutrino neutral current DIS. There is no reduction of the
extracted structure function in the neutrino charged current DIS.  The
results for $F_2$ extracted from the CCFR neutrino DIS experiment and
the ones obtained from various charged lepton neutral current DIS
experiment do indeed show such a tendency.

\item
The violation of the charge symmetry in the unpolarized structure
function $F_2$ is due to non-linear effects. The violation is derived based on
the assumption that the inclusive cross section at small $x$ can be
written as
\begin{equation}
     F_2(x,Q^2) = \sum_i e_i^2 f_i(x,Q^2),
\end{equation}
where the summation is over all possible flavors and $f_i(x,Q^2)$
is {\em independent of the charge $e_i^2$}.

This assumption is true in the normal cases since the fine structure
constant $\alpha_{em}$ for EM interaction is less than $1\%$; higher
order EM correction to $F_2$ can be neglected.

It is not true in case that the EM gauge symmetry is partially broken
for a nucleon. Take Fig.\ref{Fig:MultiPart}.b for example, the EM
induced high order {\em effective} EM interaction between the upper
and lower blocks can not be ignored. This is because the propagator
$G_T^{\mu\nu}$ contains the strong interaction component
$-G_L^{\mu\nu}$ (see Eq. \ref{GTstrng}). According to
Ref. \cite{cspc1}, $G_L^{\mu\nu} \sim 1/m_\gamma^2 \sim e_S^{-2}$ with
$e_S$ the broken charge\footnote{ The isoscalar component of the up
and down quark charge at low density.}.  But Eqs. \ref{Tfi1} and
\ref{Semi-lp-T} tell us that the coupling mediated by $G_T^{\mu\nu}$
is proportional to the product of the core part of the whole EM
charges of the two interacting quarks inside the nucleon. So the
strength of the force mediated by $G_T^{\mu\nu}$ in this case is in
the strong interaction range.  Higher order effects are important.
The magic of the partial breaking of the EM gauge symmetry inside a
nucleon is to make $f_i(x,Q^2)$ at small $x$ to depend on the relative
charge $\tilde e_i/e_S$ due to the unscreened contribution of the
 Goldstone boson, namely
\begin{equation}
    f_i = f_i(x,Q^2;r_i),
\end{equation}
with $r_i = \tilde e_i/e_S$ and $\tilde e_i$ the core part of the quark's
EM charge.

If rigorous charge symmetry condition 
\begin{equation}
  f_{u/d}^P(x,Q^2;r)-f_{d/u}^N(x,Q^2;r) = 0\label{CSym}
\end{equation}
is assumed, where the ``$P$'' denotes a proton and ``$N$'' denotes a
neutron, then the compared quantities in the literature, which are
\begin{eqnarray}
  \Delta_1 &=& f_u^P(x,Q^2;r_u)-f_d^N(x,Q^2;r_d),\\
  \Delta_2 &=& f_d^P(x,Q^2;r_d)-f_u^N(x,Q^2;r_u),
\end{eqnarray}
are not zero if the EM gauge symmetry is partially broken inside an
nucleon. This is due to the fact that the compared quantities have
different arguments.

So the finding of Refs. \cite{ChargeS} and \cite{ChargeS2} can be
interpreted as an indication of the partial breaking of the 
EM gauge symmetry inside a nucleon provided that 
Eq. \ref{CSym} is true.

\item
 
The polarized structure functions $g_1$ and $g_2$ can not be extracted
from the DIS data even in principle if the symmetry breaking term
$g_c\ne 0$ since it is $\tilde g_1$ and $\tilde g_2$ (see Eqs.
\ref{tildeg1} and \ref{tildeg2}) that are related to the experimental
observables $A_1$ and $A_2$. The extracted values for $g_1$ and $g_2$
also contain our uncertainty with the value of $F_1$ at small $x$ even
both $A_1$ and $A_2$ can be measured separately.  The experimentally
extracted $\tilde g_1$ contains a small singular piece $g_c \sim
x^{-\alpha_{{\cal P}'}}\sim x^{-1.4}$ if the symmetry breaking effects
are present.  It is possible that the rapid decrease of $g_1$ for a
neutron reported in Ref. \cite{G1N} and the rapid increase of the same
quantity for a proton implied in the original data of Ref. \cite{SMC}
are manifestations of a finite $g_c$ for both a proton and a neutron.
The most recent measurement of $A_1$ at SMC for $x$ as low as
$10^{-4}$ for a proton is plotted in Fig.  \ref{Fig:SMC} without the
errors been displayed. The solid line that fits the data has a power
law behavior of
\begin{equation}
        A_1^P \approx 1.94\times 10^{-3} x^{-0.23}.
\end{equation}
Although the power $-0.23$ may not be taken seriously at the present
stage of the experimental accuracy, one message is quite certain from
that data: $A_1$ does not show a trend that approach to zero when
$x\ge 10^{-4}$. Given $F_1\sim F_2/x$ as a working hypothesis, it
leads to $g_1^P \sim x^{-1.6}$, which is quite singular.
\begin{figure}[h]
\setlength{\unitlength}{0.240900pt}
\ifx\plotpoint\undefined\newsavebox{\plotpoint}\fi
\begin{picture}(1200,720)(0,0)
\font\gnuplot=cmr10 at 10pt
\gnuplot
\sbox{\plotpoint}{\rule[-0.200pt]{0.400pt}{0.400pt}}%
\put(370.0,217.0){\rule[-0.200pt]{4.818pt}{0.400pt}}
\put(350,217){\makebox(0,0)[r]{0}}
\put(1450.0,217.0){\rule[-0.200pt]{4.818pt}{0.400pt}}
\put(370.0,334.0){\rule[-0.200pt]{4.818pt}{0.400pt}}
\put(350,334){\makebox(0,0)[r]{40}}
\put(1450.0,334.0){\rule[-0.200pt]{4.818pt}{0.400pt}}
\put(370.0,452.0){\rule[-0.200pt]{4.818pt}{0.400pt}}
\put(350,452){\makebox(0,0)[r]{80}}
\put(1450.0,452.0){\rule[-0.200pt]{4.818pt}{0.400pt}}
\put(370.0,569.0){\rule[-0.200pt]{4.818pt}{0.400pt}}
\put(350,569){\makebox(0,0)[r]{120}}
\put(1450.0,569.0){\rule[-0.200pt]{4.818pt}{0.400pt}}
\put(370.0,687.0){\rule[-0.200pt]{4.818pt}{0.400pt}}
\put(350,687){\makebox(0,0)[r]{160}}
\put(1450.0,687.0){\rule[-0.200pt]{4.818pt}{0.400pt}}
\put(370.0,158.0){\rule[-0.200pt]{0.400pt}{4.818pt}}
\put(370,117){\makebox(0,0){0.0001}}
\put(370.0,696.0){\rule[-0.200pt]{0.400pt}{4.818pt}}
\put(474.0,158.0){\rule[-0.200pt]{0.400pt}{2.409pt}}
\put(474.0,706.0){\rule[-0.200pt]{0.400pt}{2.409pt}}
\put(535.0,158.0){\rule[-0.200pt]{0.400pt}{2.409pt}}
\put(535.0,706.0){\rule[-0.200pt]{0.400pt}{2.409pt}}
\put(579.0,158.0){\rule[-0.200pt]{0.400pt}{2.409pt}}
\put(579.0,706.0){\rule[-0.200pt]{0.400pt}{2.409pt}}
\put(612.0,158.0){\rule[-0.200pt]{0.400pt}{2.409pt}}
\put(612.0,706.0){\rule[-0.200pt]{0.400pt}{2.409pt}}
\put(640.0,158.0){\rule[-0.200pt]{0.400pt}{2.409pt}}
\put(640.0,706.0){\rule[-0.200pt]{0.400pt}{2.409pt}}
\put(663.0,158.0){\rule[-0.200pt]{0.400pt}{2.409pt}}
\put(663.0,706.0){\rule[-0.200pt]{0.400pt}{2.409pt}}
\put(683.0,158.0){\rule[-0.200pt]{0.400pt}{2.409pt}}
\put(683.0,706.0){\rule[-0.200pt]{0.400pt}{2.409pt}}
\put(700.0,158.0){\rule[-0.200pt]{0.400pt}{2.409pt}}
\put(700.0,706.0){\rule[-0.200pt]{0.400pt}{2.409pt}}
\put(716.0,158.0){\rule[-0.200pt]{0.400pt}{4.818pt}}
\put(716,117){\makebox(0,0){0.001}}
\put(716.0,696.0){\rule[-0.200pt]{0.400pt}{4.818pt}}
\put(821.0,158.0){\rule[-0.200pt]{0.400pt}{2.409pt}}
\put(821.0,706.0){\rule[-0.200pt]{0.400pt}{2.409pt}}
\put(882.0,158.0){\rule[-0.200pt]{0.400pt}{2.409pt}}
\put(882.0,706.0){\rule[-0.200pt]{0.400pt}{2.409pt}}
\put(925.0,158.0){\rule[-0.200pt]{0.400pt}{2.409pt}}
\put(925.0,706.0){\rule[-0.200pt]{0.400pt}{2.409pt}}
\put(958.0,158.0){\rule[-0.200pt]{0.400pt}{2.409pt}}
\put(958.0,706.0){\rule[-0.200pt]{0.400pt}{2.409pt}}
\put(986.0,158.0){\rule[-0.200pt]{0.400pt}{2.409pt}}
\put(986.0,706.0){\rule[-0.200pt]{0.400pt}{2.409pt}}
\put(1009.0,158.0){\rule[-0.200pt]{0.400pt}{2.409pt}}
\put(1009.0,706.0){\rule[-0.200pt]{0.400pt}{2.409pt}}
\put(1029.0,158.0){\rule[-0.200pt]{0.400pt}{2.409pt}}
\put(1029.0,706.0){\rule[-0.200pt]{0.400pt}{2.409pt}}
\put(1047.0,158.0){\rule[-0.200pt]{0.400pt}{2.409pt}}
\put(1047.0,706.0){\rule[-0.200pt]{0.400pt}{2.409pt}}
\put(1063.0,158.0){\rule[-0.200pt]{0.400pt}{4.818pt}}
\put(1063,117){\makebox(0,0){0.01}}
\put(1063.0,696.0){\rule[-0.200pt]{0.400pt}{4.818pt}}
\put(1167.0,158.0){\rule[-0.200pt]{0.400pt}{2.409pt}}
\put(1167.0,706.0){\rule[-0.200pt]{0.400pt}{2.409pt}}
\put(1228.0,158.0){\rule[-0.200pt]{0.400pt}{2.409pt}}
\put(1228.0,706.0){\rule[-0.200pt]{0.400pt}{2.409pt}}
\put(1271.0,158.0){\rule[-0.200pt]{0.400pt}{2.409pt}}
\put(1271.0,706.0){\rule[-0.200pt]{0.400pt}{2.409pt}}
\put(1305.0,158.0){\rule[-0.200pt]{0.400pt}{2.409pt}}
\put(1305.0,706.0){\rule[-0.200pt]{0.400pt}{2.409pt}}
\put(1332.0,158.0){\rule[-0.200pt]{0.400pt}{2.409pt}}
\put(1332.0,706.0){\rule[-0.200pt]{0.400pt}{2.409pt}}
\put(1355.0,158.0){\rule[-0.200pt]{0.400pt}{2.409pt}}
\put(1355.0,706.0){\rule[-0.200pt]{0.400pt}{2.409pt}}
\put(1375.0,158.0){\rule[-0.200pt]{0.400pt}{2.409pt}}
\put(1375.0,706.0){\rule[-0.200pt]{0.400pt}{2.409pt}}
\put(1393.0,158.0){\rule[-0.200pt]{0.400pt}{2.409pt}}
\put(1393.0,706.0){\rule[-0.200pt]{0.400pt}{2.409pt}}
\put(1409.0,158.0){\rule[-0.200pt]{0.400pt}{4.818pt}}
\put(1409,117){\makebox(0,0){0.1}}
\put(1409.0,696.0){\rule[-0.200pt]{0.400pt}{4.818pt}}
\put(370.0,158.0){\rule[-0.200pt]{264.990pt}{0.400pt}}
\put(1470.0,158.0){\rule[-0.200pt]{0.400pt}{134.422pt}}
\put(370.0,716.0){\rule[-0.200pt]{264.990pt}{0.400pt}}
\put(920,76){\makebox(0,0){$x$}}
\put(161,437){\makebox(0,0)[l]{$x^{-1}A_1^P$}}
\put(920,521){\makebox(0,0)[l]{$A_1^P \approx 0.00194\times x^{-0.23}$}}
\put(370.0,158.0){\rule[-0.200pt]{0.400pt}{134.422pt}}
\put(370,687){\raisebox{-.8pt}{\makebox(0,0){$\Diamond$}}}
\put(489,417){\raisebox{-.8pt}{\makebox(0,0){$\Diamond$}}}
\put(575,232){\raisebox{-.8pt}{\makebox(0,0){$\Diamond$}}}
\put(647,198){\raisebox{-.8pt}{\makebox(0,0){$\Diamond$}}}
\put(716,278){\raisebox{-.8pt}{\makebox(0,0){$\Diamond$}}}
\put(787,299){\raisebox{-.8pt}{\makebox(0,0){$\Diamond$}}}
\put(854,238){\raisebox{-.8pt}{\makebox(0,0){$\Diamond$}}}
\put(936,212){\raisebox{-.8pt}{\makebox(0,0){$\Diamond$}}}
\put(1025,227){\raisebox{-.8pt}{\makebox(0,0){$\Diamond$}}}
\put(1116,235){\raisebox{-.8pt}{\makebox(0,0){$\Diamond$}}}
\put(1197,225){\raisebox{-.8pt}{\makebox(0,0){$\Diamond$}}}
\put(1249,226){\raisebox{-.8pt}{\makebox(0,0){$\Diamond$}}}
\put(1301,218){\raisebox{-.8pt}{\makebox(0,0){$\Diamond$}}}
\put(1370,226){\raisebox{-.8pt}{\makebox(0,0){$\Diamond$}}}
\put(1438,225){\raisebox{-.8pt}{\makebox(0,0){$\Diamond$}}}
\put(370,686){\usebox{\plotpoint}}
\multiput(370.58,679.40)(0.492,-1.911){19}{\rule{0.118pt}{1.591pt}}
\multiput(369.17,682.70)(11.000,-37.698){2}{\rule{0.400pt}{0.795pt}}
\multiput(381.58,639.00)(0.492,-1.722){19}{\rule{0.118pt}{1.445pt}}
\multiput(380.17,642.00)(11.000,-34.000){2}{\rule{0.400pt}{0.723pt}}
\multiput(392.58,602.45)(0.492,-1.581){19}{\rule{0.118pt}{1.336pt}}
\multiput(391.17,605.23)(11.000,-31.226){2}{\rule{0.400pt}{0.668pt}}
\multiput(403.58,568.91)(0.492,-1.439){19}{\rule{0.118pt}{1.227pt}}
\multiput(402.17,571.45)(11.000,-28.453){2}{\rule{0.400pt}{0.614pt}}
\multiput(414.58,538.71)(0.492,-1.186){21}{\rule{0.119pt}{1.033pt}}
\multiput(413.17,540.86)(12.000,-25.855){2}{\rule{0.400pt}{0.517pt}}
\multiput(426.58,510.66)(0.492,-1.203){19}{\rule{0.118pt}{1.045pt}}
\multiput(425.17,512.83)(11.000,-23.830){2}{\rule{0.400pt}{0.523pt}}
\multiput(437.58,484.96)(0.492,-1.109){19}{\rule{0.118pt}{0.973pt}}
\multiput(436.17,486.98)(11.000,-21.981){2}{\rule{0.400pt}{0.486pt}}
\multiput(448.58,461.41)(0.492,-0.967){19}{\rule{0.118pt}{0.864pt}}
\multiput(447.17,463.21)(11.000,-19.207){2}{\rule{0.400pt}{0.432pt}}
\multiput(459.58,440.57)(0.492,-0.920){19}{\rule{0.118pt}{0.827pt}}
\multiput(458.17,442.28)(11.000,-18.283){2}{\rule{0.400pt}{0.414pt}}
\multiput(470.58,420.87)(0.492,-0.826){19}{\rule{0.118pt}{0.755pt}}
\multiput(469.17,422.43)(11.000,-16.434){2}{\rule{0.400pt}{0.377pt}}
\multiput(481.58,403.17)(0.492,-0.732){19}{\rule{0.118pt}{0.682pt}}
\multiput(480.17,404.58)(11.000,-14.585){2}{\rule{0.400pt}{0.341pt}}
\multiput(492.58,387.32)(0.492,-0.684){19}{\rule{0.118pt}{0.645pt}}
\multiput(491.17,388.66)(11.000,-13.660){2}{\rule{0.400pt}{0.323pt}}
\multiput(503.58,372.47)(0.492,-0.637){19}{\rule{0.118pt}{0.609pt}}
\multiput(502.17,373.74)(11.000,-12.736){2}{\rule{0.400pt}{0.305pt}}
\multiput(514.58,358.79)(0.492,-0.539){21}{\rule{0.119pt}{0.533pt}}
\multiput(513.17,359.89)(12.000,-11.893){2}{\rule{0.400pt}{0.267pt}}
\multiput(526.00,346.92)(0.496,-0.492){19}{\rule{0.500pt}{0.118pt}}
\multiput(526.00,347.17)(9.962,-11.000){2}{\rule{0.250pt}{0.400pt}}
\multiput(537.00,335.92)(0.496,-0.492){19}{\rule{0.500pt}{0.118pt}}
\multiput(537.00,336.17)(9.962,-11.000){2}{\rule{0.250pt}{0.400pt}}
\multiput(548.00,324.93)(0.611,-0.489){15}{\rule{0.589pt}{0.118pt}}
\multiput(548.00,325.17)(9.778,-9.000){2}{\rule{0.294pt}{0.400pt}}
\multiput(559.00,315.93)(0.611,-0.489){15}{\rule{0.589pt}{0.118pt}}
\multiput(559.00,316.17)(9.778,-9.000){2}{\rule{0.294pt}{0.400pt}}
\multiput(570.00,306.93)(0.692,-0.488){13}{\rule{0.650pt}{0.117pt}}
\multiput(570.00,307.17)(9.651,-8.000){2}{\rule{0.325pt}{0.400pt}}
\multiput(581.00,298.93)(0.798,-0.485){11}{\rule{0.729pt}{0.117pt}}
\multiput(581.00,299.17)(9.488,-7.000){2}{\rule{0.364pt}{0.400pt}}
\multiput(592.00,291.93)(0.798,-0.485){11}{\rule{0.729pt}{0.117pt}}
\multiput(592.00,292.17)(9.488,-7.000){2}{\rule{0.364pt}{0.400pt}}
\multiput(603.00,284.93)(0.943,-0.482){9}{\rule{0.833pt}{0.116pt}}
\multiput(603.00,285.17)(9.270,-6.000){2}{\rule{0.417pt}{0.400pt}}
\multiput(614.00,278.93)(1.267,-0.477){7}{\rule{1.060pt}{0.115pt}}
\multiput(614.00,279.17)(9.800,-5.000){2}{\rule{0.530pt}{0.400pt}}
\multiput(626.00,273.93)(1.155,-0.477){7}{\rule{0.980pt}{0.115pt}}
\multiput(626.00,274.17)(8.966,-5.000){2}{\rule{0.490pt}{0.400pt}}
\multiput(637.00,268.93)(1.155,-0.477){7}{\rule{0.980pt}{0.115pt}}
\multiput(637.00,269.17)(8.966,-5.000){2}{\rule{0.490pt}{0.400pt}}
\multiput(648.00,263.94)(1.505,-0.468){5}{\rule{1.200pt}{0.113pt}}
\multiput(648.00,264.17)(8.509,-4.000){2}{\rule{0.600pt}{0.400pt}}
\multiput(659.00,259.94)(1.505,-0.468){5}{\rule{1.200pt}{0.113pt}}
\multiput(659.00,260.17)(8.509,-4.000){2}{\rule{0.600pt}{0.400pt}}
\multiput(670.00,255.95)(2.248,-0.447){3}{\rule{1.567pt}{0.108pt}}
\multiput(670.00,256.17)(7.748,-3.000){2}{\rule{0.783pt}{0.400pt}}
\multiput(681.00,252.94)(1.505,-0.468){5}{\rule{1.200pt}{0.113pt}}
\multiput(681.00,253.17)(8.509,-4.000){2}{\rule{0.600pt}{0.400pt}}
\put(692,248.17){\rule{2.300pt}{0.400pt}}
\multiput(692.00,249.17)(6.226,-2.000){2}{\rule{1.150pt}{0.400pt}}
\multiput(703.00,246.95)(2.248,-0.447){3}{\rule{1.567pt}{0.108pt}}
\multiput(703.00,247.17)(7.748,-3.000){2}{\rule{0.783pt}{0.400pt}}
\multiput(714.00,243.95)(2.472,-0.447){3}{\rule{1.700pt}{0.108pt}}
\multiput(714.00,244.17)(8.472,-3.000){2}{\rule{0.850pt}{0.400pt}}
\put(726,240.17){\rule{2.300pt}{0.400pt}}
\multiput(726.00,241.17)(6.226,-2.000){2}{\rule{1.150pt}{0.400pt}}
\put(737,238.17){\rule{2.300pt}{0.400pt}}
\multiput(737.00,239.17)(6.226,-2.000){2}{\rule{1.150pt}{0.400pt}}
\put(748,236.17){\rule{2.300pt}{0.400pt}}
\multiput(748.00,237.17)(6.226,-2.000){2}{\rule{1.150pt}{0.400pt}}
\put(759,234.67){\rule{2.650pt}{0.400pt}}
\multiput(759.00,235.17)(5.500,-1.000){2}{\rule{1.325pt}{0.400pt}}
\put(770,233.17){\rule{2.300pt}{0.400pt}}
\multiput(770.00,234.17)(6.226,-2.000){2}{\rule{1.150pt}{0.400pt}}
\put(781,231.67){\rule{2.650pt}{0.400pt}}
\multiput(781.00,232.17)(5.500,-1.000){2}{\rule{1.325pt}{0.400pt}}
\put(792,230.17){\rule{2.300pt}{0.400pt}}
\multiput(792.00,231.17)(6.226,-2.000){2}{\rule{1.150pt}{0.400pt}}
\put(803,228.67){\rule{2.650pt}{0.400pt}}
\multiput(803.00,229.17)(5.500,-1.000){2}{\rule{1.325pt}{0.400pt}}
\put(814,227.67){\rule{2.891pt}{0.400pt}}
\multiput(814.00,228.17)(6.000,-1.000){2}{\rule{1.445pt}{0.400pt}}
\put(826,226.67){\rule{2.650pt}{0.400pt}}
\multiput(826.00,227.17)(5.500,-1.000){2}{\rule{1.325pt}{0.400pt}}
\put(837,225.67){\rule{2.650pt}{0.400pt}}
\multiput(837.00,226.17)(5.500,-1.000){2}{\rule{1.325pt}{0.400pt}}
\put(848,224.67){\rule{2.650pt}{0.400pt}}
\multiput(848.00,225.17)(5.500,-1.000){2}{\rule{1.325pt}{0.400pt}}
\put(870,223.67){\rule{2.650pt}{0.400pt}}
\multiput(870.00,224.17)(5.500,-1.000){2}{\rule{1.325pt}{0.400pt}}
\put(881,222.67){\rule{2.650pt}{0.400pt}}
\multiput(881.00,223.17)(5.500,-1.000){2}{\rule{1.325pt}{0.400pt}}
\put(859.0,225.0){\rule[-0.200pt]{2.650pt}{0.400pt}}
\put(903,221.67){\rule{2.650pt}{0.400pt}}
\multiput(903.00,222.17)(5.500,-1.000){2}{\rule{1.325pt}{0.400pt}}
\put(892.0,223.0){\rule[-0.200pt]{2.650pt}{0.400pt}}
\put(926,220.67){\rule{2.650pt}{0.400pt}}
\multiput(926.00,221.17)(5.500,-1.000){2}{\rule{1.325pt}{0.400pt}}
\put(914.0,222.0){\rule[-0.200pt]{2.891pt}{0.400pt}}
\put(959,219.67){\rule{2.650pt}{0.400pt}}
\multiput(959.00,220.17)(5.500,-1.000){2}{\rule{1.325pt}{0.400pt}}
\put(937.0,221.0){\rule[-0.200pt]{5.300pt}{0.400pt}}
\put(992,218.67){\rule{2.650pt}{0.400pt}}
\multiput(992.00,219.17)(5.500,-1.000){2}{\rule{1.325pt}{0.400pt}}
\put(970.0,220.0){\rule[-0.200pt]{5.300pt}{0.400pt}}
\put(1048,217.67){\rule{2.650pt}{0.400pt}}
\multiput(1048.00,218.17)(5.500,-1.000){2}{\rule{1.325pt}{0.400pt}}
\put(1003.0,219.0){\rule[-0.200pt]{10.840pt}{0.400pt}}
\put(1148,216.67){\rule{2.650pt}{0.400pt}}
\multiput(1148.00,217.17)(5.500,-1.000){2}{\rule{1.325pt}{0.400pt}}
\put(1059.0,218.0){\rule[-0.200pt]{21.440pt}{0.400pt}}
\put(1159.0,217.0){\rule[-0.200pt]{74.920pt}{0.400pt}}
\end{picture}
\caption{\label{Fig:SMC} The value of $A_1$ for a proton measured in
the most recent SMC publication.  It is displayed in such a way that
its tendency of divergence at small $x$ is more transparent. The solid
line is a power law fit to the somewhat oscillating data points. The
extracted structure function $\widetilde g_1^P$ is proportional to
$F^P_2 A^P_1/x$.  }
\end{figure}
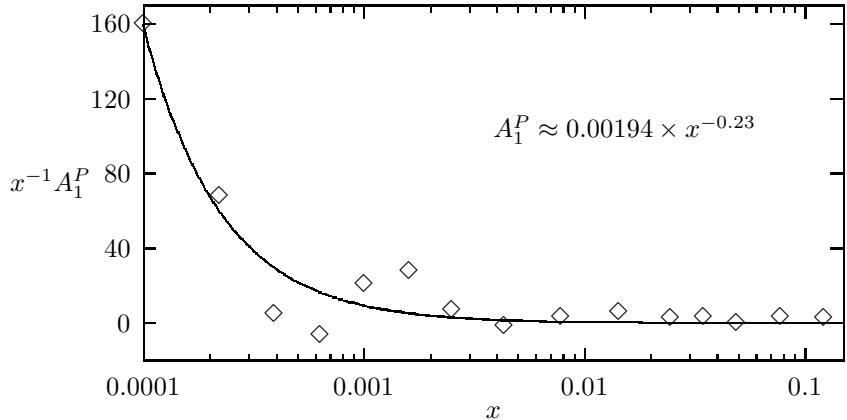

However such a singular behavior for $g_1$ is of high twist effects,
it is expected to go a way in the $\nu\to \infty$ limit.  The
contribution of $g_c$ to $\widetilde g_2$ is not suppressed in the
Bjorken limit. In addition it is more singular than $\widetilde g_1$
since $\widetilde g_2 \sim g_c/x$.

\item

The spin ``crisis'' is still with us. It can mean either 1) the
sea quark of a nucleon is negatively polarized against the nucleon
spin or/and 2) the orbital angular momentum of the quarks is finite or/and
3) the contribution of the gluons polarization through the anomalous term
or/and 4) the strange content of a nucleon is finite to
provide the rest of the first moment $\Gamma_1$.  The recent
measurement of the sea quark polarization turns out to be quite small
\cite{SeaPol}. However there is still a lack of the gluons polarization
$\Delta G$ data at the present. It is possible that such a picture is
correct for $g_1$. But the recently found rapid change of $g_1(x)$ at
small $x$ could make the crisis even more severe.

The assumption of a partial spontaneous breaking of the EM gauge
symmetry could provide at least a partial understanding for the spin
crisis related problems. This is because the extracted $\tilde g_1$
contains a singular piece $g_c\sim x^{-\alpha_{{\cal P}'}}$,
which should give a divergent result for the first moment
$\Gamma_1$ as the lower bound of the $x$ integration is getting
smaller and smaller in future measurements. 
The reason the present
value of $\Gamma_1$ to be finite is due to the assumed normal Regge
behavior extrapolation from the known data at larger $x$.

\item

The Bjorken sum rule should be respected at least in the Bjorken
limit.  It is not clear whether or not the Bjorken sum rule is
violated by the experimental data at finite $\nu$ and $Q^2$ due to the
presence of $g_c$ in the extracted $\tilde g_1$ in the case of
symmetry breaking when more data for $\tilde g_1$ at smaller $x$ is
known. The reason that the Bjorken sum rule appears to be not violated
in the current experimental data could mean either that it is in fact
not violated or it is not violated if a normal Regge behavior for
$g_1$ at small $x$ is assumed. In either case, it only mean that
the Bjorken sum rule is respected by the regular $g_1$.

It is highly possible the the Bjorken sum rule is violated by $\tilde
g_1$ if more experimental information about the its small $x$ behavior
is known.  In that case, the isospin odd component of the imaginary
part of $c$ or $Z_c$ is different from zero, which is allowed under
the scenario proposed here (see Eq. \ref{EQ3}). But it must be
emphasized that the possible violate of the isospin odd component of
the GDH sum rule and the possible violation of the Bjorken sum rule is
not in one to one correspondence even in symmetry breaking case.  The
violation of the GDH sum rule is due to the asymptotic polynomial part
of $c$ defined in Eq. \ref{c-disp}, which does not contribute to
quantities involved in the DIS cross sections.  It may or may not be
zero even in the symmetry breaking case.

\item

For the same reason as the ones given above, the Burkhardt--Cottingham
sum rule is likely to be violated when more data about $A_2$ (or
$\widetilde g_2$) is known at small $x$ due to the singular piece
proportional to $g_c$. It will remain to be so even in the true
Bjorken limit (see Eq. \ref{g2-Blmt}) if the nucleon is
superconducting.

\item

The energy dependence of the $\rho$ electroproduction follows that of
the soft pomeron dominance in the observation. This is expected even
in the symmetry breaking case. The rapid increase of the $\phi$ and
$J/\Psi$ production cross section \cite{JPsidata,data2} that is dominated by
the hard pomeron can be incorporated in the current scenario if the
partial EM gauge symmetry breaking for a nucleon is assumed. This
is due to the fact that like a lepton, the strange and charm quarks
contain no nucleon number. They do not couple to the Goldstone boson
from EM $U(1)$ symmetry breaking. The effective EM coupling of strange
and charm quark to the up and down quark sector contains a strong
interaction component.

\item 
 
The current experimental data on the $pp$ collision production of
mesons seems to indicate \cite{Close} that the soft pomeron transforms
as a vector that couples to the a non-conserving current. This is a
natural consequences of the scenario proposed here if one assumes that
there is a spontaneous EM gauge symmetry partial breaking phase for
a nucleon.  This is because if the soft pomeron behaves like a photon,
then only the non-conserving core component of the EM charge current
in the color superconducting phase contribute to the observables, as
it is discussed above.

\item

 The scenario proposed in this work depends on the assumption that
there is a nucleon long before the experimental processes take place.
For those $N\bar N$ creation processes like in the $\gamma\gamma$
collision in the vacuum, the time for the initial $Q\overline Q$ 
creation is expected to be
much faster than the collective multi-particle processes to let the
excited system to  cool to its ground state containing the observed
$N\bar N$ pairs
, in which the possible superconducting
phase is formed. So, the proposed hard pomeron behavior should not
appear \cite{PCAC}. This is indeed observed behavior \cite{gamma2}.

\end{enumerate}

\section{Summary}
\label{sec:summary}

    The possibility to search for a close-by metastable color
superconducting phase for the strong interaction vacuum state in the
presence of a nucleon is studied. It is shown that the metastable
color superconducting phase of the strong interaction vacuum state can
manifest itself in high energy semi-leptonic neutral current
electroweak interaction processes involving a nucleon through a
mechanism of ``spontaneous partial breaking'' of the EM $U_{em}(1)$
gauge symmetry that is independent of any mass scale.

    In order to achieve this, the physical processes in a system in
which the EM gauge symmetry is spontaneously partial broken leading to
color superconductivity is studied model independently. It is found
that due the fact that the electric charge of quarks is fractional of
the charge of a proton and is flavor dependent, the spontaneous
partial breaking of the EM gauge symmetry in the hadronic sector
caused by a diquark condensation also breaks the global baryon number
conservation.  The EM gauge symmetry is only spontaneously partial
broken in such a case where the Goldstone bosons corresponding to the
global symmetry breaking is not unphysical states in all channels of
the reaction.  It is also shown that for a system in which the EM
gauge symmetry is spontaneously partial broken, due to the medium
effects, the final phase space in a high energy semi-leptonic process
contains an extra subspace compared to the allowed one due to the
Froissart bound for the high energy nucleon--nucleon scattering. Such
an extra subspace can be considered as spanned by the (lack of)
Goldstone bosons of the the spontaneous global symmetry breaking which
induces the spontaneous partial breaking of the EM gauge symmetry.

     It is found here that to explain some of the puzzling behavior of
the nucleon structure functions at small $x$, the violation of the GDH
sum rule, the meson production in exclusive semi-leptonic DIS and the
current non-conservation in the meson production in high energy
proton--proton collisions, the assumption that a nucleon is
superconducting needs to be made.  The spin structure functions of the
nucleons at small $x$ is perhaps the most sensitive ones to study the
spontaneous partial breaking of EM gauge symmetry inside a nucleon due
to the presence of a very singular component in the experimental
observables $\widetilde g_1(x)$ and $\widetilde g_2(x)$, which was
seen in the most recent data at $x$ as low as $10^{-4}$ for $\tilde
g_1(x)$. The magnitude of the effects of the spontaneous partial
breaking of the EM gauge symmetry observed in the EM properties of a
nucleon at high energies, if proven true, means that there is at least
one color superconducting metastable phase for the hadronic vacuum
state with low enough energy density\footnote{The energy density of
the true vacuum state is assumed to be zero.}  that can be turned into
the stable one at relatively low density (compared to the nuclear
saturation density) rather than at high densities.  Because of the
simplicity of the single nucleon system, the signal for the possible
color superconducting phase of the vacuum state is relatively clear to
allow a further more detailed experimental study in the existing and
planed facilities measuring the high energy EM and neutral current
weak responses of a nucleon at small $Q^2$ and at small $x$.

   It must be emphasized that the results of this paper and earlier
related ones, which concern with the virtual phases of the vacuum
state, does neither uniquely imply nor disfavor the popular diquark
model or quark--quark clustering model for a nucleon. These models are
based on hypothetical binding/clustering between a pair of valence
quarks inside a nucleon. Such a binding/clustering at valence level
has no direct relation to the conclusion draw here. In fact a model
for a nucleon can be build without any quark--quark clustering even in
the presence of a virtual color superconducting phase for the vacuum
\cite{model}.

\section*{Acknowledgements}

  This work is supported by the National Natural Science Foundation of
China under contract 19875009  and a research fund from the 
Department of Education of China.

\end{document}